\documentclass[twocolumn,trackchanges]{aastex631}
\usepackage{amsmath}
\usepackage{epsf}
\usepackage{color}
\usepackage{color}
\usepackage{enumitem}
\usepackage{mathtools}
\usepackage[mathcal]{eucal}
 \let\mathscr\relax
\usepackage[scr]{rsfso}

\shorttitle{Co-Evolution of the Rest-UV Luminosity Function}
\shortauthors{Finkelstein \& Bagley}

\newcommand{\sol}{$_{\odot}$}

\newcommand{\muv}{$M_\mathrm{UV}$}

\begin{document}
\title{On the Co-Evolution of the AGN and Star-Forming Galaxy Ultraviolet Luminosity Functions at 3 $< z <$ 9}
\author[0000-0001-8519-1130]{Steven L. Finkelstein}
\affiliation{Department of Astronomy, The University of Texas at Austin, Austin, TX, USA}
\email{stevenf@astro.as.utexas.edu}

\author[0000-0002-9921-9218]{Micaela B. Bagley}
\affiliation{Department of Astronomy, The University of Texas at Austin, Austin, TX, USA}

\begin{abstract}
Studies of the high-redshift rest-frame ultraviolet (UV) luminosity functions have typically treated the star-forming galaxy and active galactic nuclei (AGN) populations separately, as they have different survey depth and area requirements.  However, the recent advent of wide-area modestly deep ground-based imaging surveys now probe volumes large enough to discover AGNs, at depths sensitive enough for fainter star-forming galaxies, bridging these two populations.  Using results from such surveys as observational constraints, we present a methodology to jointly empirically model the evolution of the rest-UV luminosity functions at $z =$ 3--9.  We assume that both populations have a luminosity function well-described by a double power law modified to allow a flattening at the faint-end, and that all luminosity function parameters evolve smoothly with redshift.  This methodology provides a good fit to the observations, and makes predictions to volume densities and redshifts not yet observed.  We find that the volume density of bright ($M_{UV} = -$28) AGNs rises by five orders of magnitude from $z =$ 9 to $z =$ 3, while modestly bright ($M_{UV} = -$21) star-forming galaxies rise by only two orders of magnitude across the same epoch. The observed bright-end flattening of the $z =$ 9 luminosity function is unlikely to be due to AGN activity, and rather is due to a shallowing of the bright-end slope, implying a reduction of feedback in bright galaxies at early times.  Integrating our luminosity functions we find that the intrinsic ionizing emissivity is dominated by star-forming galaxies at all redshifts explored, and this result holds even after applying a notional escape fraction to both populations.  We compare our AGN luminosity functions to predicted AGN luminosity functions based on varied black-hole seeding models, finding decent agreement on average, but that all models are unable to predict the observed abundance of bright AGNs.  We make predictions for the upcoming {\it Euclid} and {\it Roman} observatories, showing that their respective wide-area surveys should be capable of discovering AGNs to $z \sim$ 8.
\end{abstract}



\section{Introduction}\label{sec:intro}

The observed ultraviolet (UV) luminosities of galaxies provide key insight into the rates of star formation and super-massive black hole (SMBH) growth.  Consequently, the distribution function of this quantity, the rest-frame UV luminosity function, is a key observable which places strong constraints on the evolution of galaxies \citep[see reviews by, e.g.,][and references therein]{somerville15,finkelstein16,stark16,robertson21}.  This is especially true at $z >$ 3, where the majority of known galaxies are selected via their rest-UV emission due to present observational limitations (and obscured star-formation is likely less important at $z >$ 4; e.g., \citealt{zavala21}).

To date, much of the literature on the UV luminosity function at $z >$ 3 has focused on objects where the luminous emission is dominated by massive stars, e.g., star-forming galaxies.  This is predominantly due to the volumes probed for such galaxies, which until recently has limited the discovered galaxy population to rest-frame UV absolute magnitudes ($M_\mathrm{UV}$) brighter than $-$23 due to the small volume densities of brighter sources \citep[e.g.,][]{bouwens15,finkelstein15,bowler15,adams20,bowler20}.  The majority of sources with $M_\mathrm{UV}$ $\gtrsim$ $-$23 appear to have their emission dominated by star-formation \citep[e.g.,][]{stevans18,bowler20,harikane21}.

The shape of the galaxy UV luminosity function has historically been well-described by a \citet{schechter76} function -- a power-law slope ($\alpha$) at faint luminosities, transitioning to an exponential decline at luminosities brighter than the characteristic luminosity $M^{\ast}$ (with a characteristic number density $\phi^{\ast}$).  This may not be surprising as this functional form is similar to that predicted for the halo mass function.  However, by comparing the observed luminosity functions to the predicted underlying halo mass function, it is clear that baryonic effects (primarily negative feedback due to supermassive black hole growth, supernovae, and stellar radiation) alter the shape of the galaxy luminosity function \citep[e.g.,][and references therein]{somerville15}.  
Consequently, there is no strong physical reason why the galaxy luminosity function should exhibit an exponential decline at the bright end.  

Perhaps unsurprisingly, recent studies using wider-field surveys have found that the shape of the bright end of the rest-UV luminosity function may exhibit a power-law decline rather than exponential, with the full star-forming galaxy luminosity function well-described by a double power law \citep[e.g.,][]{stevans18,ono18,stefanon19,bowler20,harikane21}.  As the shape of the UV luminosity function encodes key information on the physics dominating how galaxies evolve, understanding how this shape evolves is of key astrophysical significance.

However, the volumes probed to reach these rare bright star-forming galaxies are also sensitive to objects where the luminosity is dominated by SMBH growth, known as active galactic nuclei (AGNs).  Studies at lower-redshift have shown that the AGN rest-UV luminosity function is also well described by a double power law, with a characteristic magnitude much brighter than that of the galaxy luminosity function and a much lower characteristic number density \citep[e.g.][]{kulkarni19,shen20}.

To constrain the evolution of the bright-end of the star-forming galaxy UV luminosity function one must thus also understand the evolution of the faint-end of the AGN UV luminosity function.  Most studies have focused on either one population or the other, employing techniques to discern AGN from star-forming dominated galaxies (e.g., morphologies to isolate point sources, or spectroscopic followup; e.g., \citealt{akiyama18}).  These techniques all have their associated biases.   One way to gain a better understanding of the co-evolution of the AGN and star-forming galaxy UV luminosity functions is to study both together.

The past few years has seen a handful of large-volume modestly deep surveys capable of probing the volumes for the more rare AGNs, but at depths sensitive to more common star-forming galaxies.  For example, \citet{stevans18} used the 24 deg$^2$ {Spitzer}-HETDEX Exploratory Large Area (SHELA) survey to study the full (AGN and star-forming galaxy) $z \sim$ 4 UV luminosity function.  \citet{adams20} performed a similar analysis empowered by the 6 deg$^2$ VIDEO survey, while more recently \citet{harikane21} used the HyperSuprimeCam Subaru Strategic Program (SSP; \citealt{aihara18}) 300 deg$^2$ survey to do a similar analysis at $z =$ 4--7.  \citet{zhang21} were also able to probe both populations simultaneously at $z \sim$ 2--3 using a Ly$\alpha$-selected sample from the Hobby Eberly Telescope Dark Energy Experiment \citep{gebhardt21} unbiased spectroscopic survey.

As shown by several of these studies, it is necessary to combine these wide-field ground based surveys that can reach the brightest AGNs with results from deeper but narrower {\it Hubble Space Telescope} ({\it HST}) surveys to reach fainter galaxies.  However, even when covering a wide dynamic range in luminosity, degeneracies in the fitting can still preclude strong constraints on the luminosity function shape.  As we expect the luminosity function to evolve smoothly as a function of the scale factor, it should be possible to improve the luminosity function constraints by fitting observational data simultaneously over a wide range of redshifts, under the assumption that both AGN and star-forming galaxy luminosity function parameters evolve as smooth polynomial functions of 1$+z$ (see \citet{kulkarni19} for a similar analysis for the AGN population only).

Here we combine the results from a number of recent surveys for both AGN and star-forming galaxies to explore the co-evolution of the AGN and star-forming galaxy rest-UV luminosity functions.  We fit the co-evolution of these combined luminosity functions over the wide redshift range of 3 $< z <$ 9, relying on the available data over this wide epoch to remove degeneracies which can plague results when studying only a single redshift \citep[e.g.,][]{stevans18}.  We place a particular emphasis on understanding the shape of the luminosity function at $z =$ 8--9, where current results imply the bright end appears to be flattening.  With our results, we shed light on whether this flattening is significant, and whether we expect the luminosities of these bright $z >$ 8 galaxies to be dominated by AGN or massive stars.

\begin{figure*}[!t]
\epsscale{1.2}
\plotone{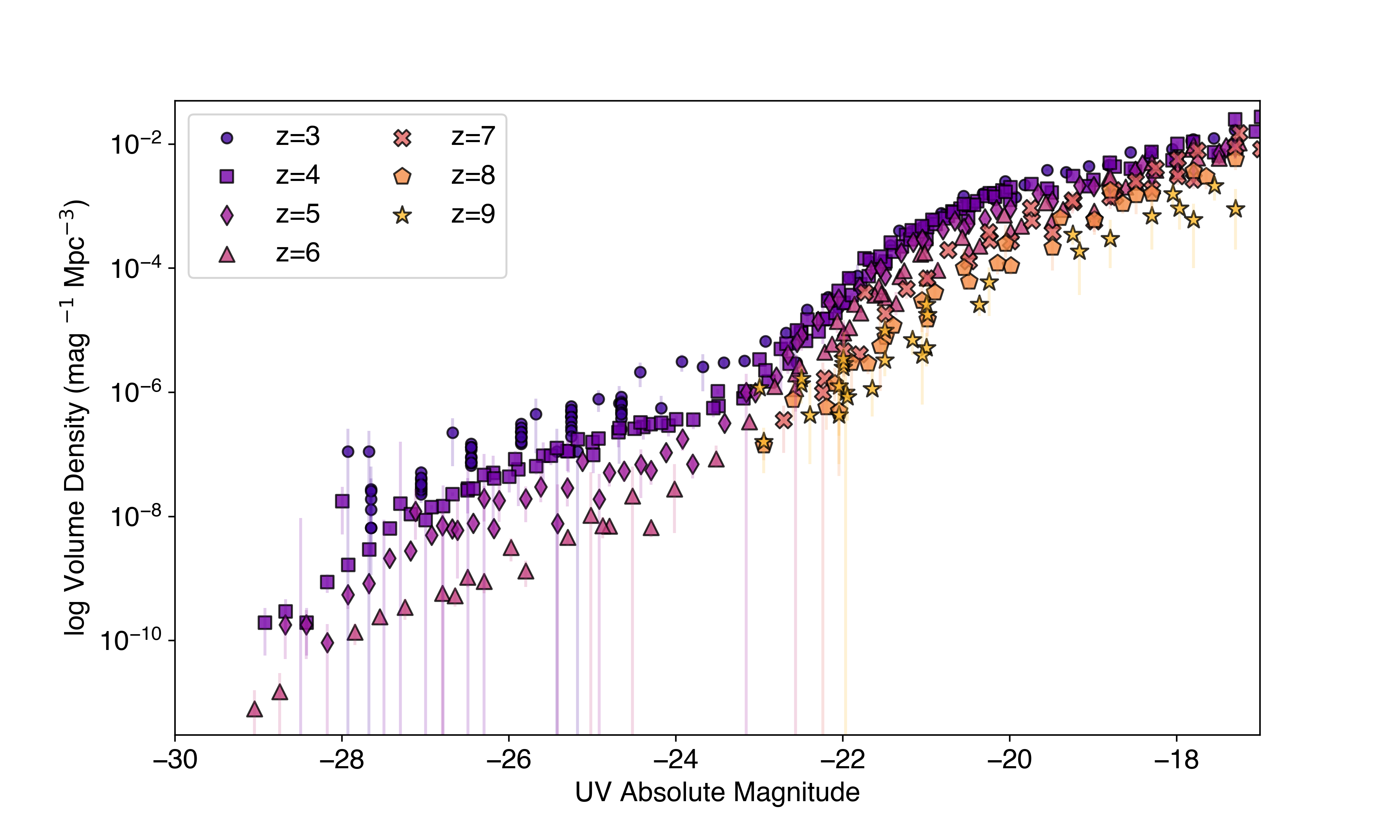}
\vspace*{-10mm}
\caption{Measurements of the rest-frame UV luminosity function taken from a variety of literature sources (described in Tables 1 and 2).  Different symbols and colors denote different redshifts, from $z =$ 3--9.  At the lowest redshifts plotted, observations span over 10 magnitudes, while the highest redshift observations are limited to the fainter half of that range.  Measurements at M $< -$ 23, probing the bulk of the AGN population, are only available at $z \leq$ 6, motivating the need to model all redshifts together to probe the evolution of both populations to $z >$ 6.}
\label{fig:fig1}
\end{figure*}

This paper is organized as follows.  In Section 2, we explain our methodology, describing the datasets used and the algorithms we developed to measure the UV luminosity functions simultaneously across multiple redshifts.  We describe our results in \S 3.  In \S 4, we discuss our results, and make predictions for future observations over large volumes with the {\it Nancy Grace Roman Space Telescope} ({\it NGRST}).  We present our conclusions in \S 5.  In this paper we assume the latest {\it Planck} flat $\Lambda$CDM cosmology with H$_{0} =$ 67.36, $\Omega_m=$0.3153, and $\Omega_{\Lambda}=$0.6847 \citep{planck20}.  All magnitudes are in the absolute bolometric system \citep[AB][]{oke83}.

\section{Methodology}\label{sec:methodology}

\subsection{Luminosity Function Data}
We use published number densities from a variety of analyses over the past several years.  While we do not make use of the entirety of the published literature, when possible we do include results from complementary studies even when published from the same imaging datasets, such that our results marginalize over differences in the analyses (e.g., different photometry methods, selection methods, completeness correction techniques).  We list the datasets used in Tables 1 (3 $\leq z \leq$ 6) and 2 (7 $\leq z \leq$ 9), and visualize the number densities across all redshifts in Figure~\ref{fig:fig1}.

\begin{deluxetable*}{ccccc}
\vspace{-2mm}
\tabletypesize{\small}
\tablecaption{Literature Data Used to Constrain Luminosity Function Evolution at $z=$3--6}
\tablewidth{\textwidth}
\tablehead{
\colhead{Redshift} & \colhead{Reference} & \colhead{M$_{UV}$ Range} & {Primary Survey$^{\dagger}$} & \colhead{Type}}
\startdata
3.0&\citet{parsa16}&[$-$22.5,$-$17.5$^{b}$]&CANDELS&Star-Forming Galaxies\\
3.0&\citet{bouwens21}&[$-$22.5,$-$17.3]&CANDELS&Star-Forming Galaxies\\
3.0&\citet{bouwens22}&[$-$18.8,$-$12.3]&Frontier Fields&Star-Forming Galaxies\\
3.0&\citet{kulkarni19}$^{a}$&[$-$27.6,$-$24.0]&SDSS&AGN\\ 
3.0&\citet{moutard20}&[$-$27.9,$-$19.9]&CLAUDS/HSC SSP&Both\\
\hline
4.0&\citet{parsa16}&[$-$22.5,$-$16.0]&CANDELS&Star-Forming Galaxies\\
4.0&\citet{finkelstein15}&[$-$22.5,$-$17.5]&CANDELS&Star-Forming Galaxies\\
4.0&\citet{bouwens21}&[$-$22.7,$-$15.9]&CANDELS&Star-Forming Galaxies\\
4.0&\citet{bouwens22}&[$-$18.8,$-$13.8]&Frontier Fields&Star-Forming Galaxies\\
4.0&\citet{akiyama18}&[$-$25.9,$-$24.1]&HSC SSP&AGN\\
4.0&\citet{akiyama18}&[$-$28.9,$-$26.1]&SDSS&AGN\\
4.0&\citet{stevans18}&[$-$28.5,$-$20.5]&SHELA&Both\\
4.0&\citet{adams20}&[$-$27.3,$-$20.1]&UltraVISTA/VIDEO&Both\\
4.0&\citet{harikane21}&[$-$26.8,$-$20.0]&HSC SSP&Both\\
\hline
5.0&\citet{finkelstein15}&[$-$22.5,$-$17.5]&CANDELS&Star-Forming Galaxies\\
5.0&\citet{bouwens21}&[$-$23.1,$-$16.4]&CANDELS&Star-Forming Galaxies\\
5.0&\citet{bouwens22}&[$-$18.8,$-$13.8]&Frontier Fields&Star-Forming Galaxies\\
5.0&\citet{mcgreer13}&[$-$28.1,$-$24.3]&CFHTLS&AGN\\
5.0&\citet{yang16}&[$-$29.0,$-$27.1]&SDSS&AGN\\
5.0&\citet{kim20}&[$-$26.8,$-$23.8]&Infrared Medium-Deep Survey&AGN\\
5.0&\citet{niida20}&[$-$27.1,$-$24.1]&HSC SSP&AGN\\
5.0&\citet{niida20}&[$-$28.6,$-$26.1]&SDSS&AGN\\
5.0&\citet{harikane21}&[$-$25.4,$-$20.3]&HSC SSP&Both\\
\hline
6.0&\citet{bowler15}&[$-$22.5,$-$21.2]&UltraVISTA/UKIDSS&Star-Forming Galaxies\\
6.0&\citet{finkelstein15}&[$-$22.0,$-$17.5]&CANDELS&Star-Forming Galaxies\\
6.0&\citet{bouwens17}&[$-$20.8,$-$12.8]&Frontier Fields&Star-Forming Galaxies\\
6.0&\citet{atek18}&[$-$21.9,$-$13.3]&Frontier Fields&Star-Forming Galaxies\\
6.0&\citet{bouwens21}&[$-$22.5,$-$16.8]&CANDELS&Star-Forming Galaxies\\
6.0&\citet{bouwens22}&[$-$18.8,$-$13.3]&Frontier Fields&Star-Forming Galaxies\\
6.0&\citet{jiang16}&[$-$29.1,$-$24.7]&SDSS&AGN\\
6.0&\citet{matsuoka18}&[$-$29.0,$-$24.3]&HSC SSP&AGN\\
6.0&\citet{harikane21}$^{c}$&[$-$25.0,$-$21.0]&HSC SSP&Both\\
\enddata
\tablecomments{The final column denotes which type of objects dominate in a given study.   Values from AGN-only studies at $M > -$24 were excluded due to potential  incompleteness from rejection of star-forming galaxies.  $^{\dagger}$This column indicates the survey which is most constraining for each study.  $^{a}$We used reported AGN number densities across 2.6 $< z <$ 3.4.  $^{b}$Did not consider absolute magnitudes within 0.5 mag of the Lyman break dropout band detection limit (which is the ground-based U-band image for Parsa et al.\ (2016) at $z =$ 3).  $^{c}$We do not use the Harikane et al.\ (2021) data at $z \sim$ 7, as these galaxies are one-band detections in the HSC SSP dataset, and are thus less certain than the other $z =$ 7 data used. 
}
\vspace{-5mm}
\label{tab:data1}
\end{deluxetable*}

The majority of studies we included select primarily star-forming galaxies (usually limited to \muv\ $> -$ 23), or quasars/AGN (primarily at brighter magnitudes).  A few more recent studies \citep{stevans18,moutard20,adams20,harikane21} probe volumes large enough to include both types of sources, which are especially important for bridging the two populations.  For the most part, we use all values published in a given paper (and/or shared via private communication).  There are two exceptions.  First, for papers which select AGNs only, we exclude all results at \muv\ $> -$24.  At fainter magnitudes star-forming dominated systems begin to contribute to the observed number densities (see \S 3), thus the AGN-only results do not represent the total population of astrophysical objects at those magnitudes.  On the galaxy side, we typically use all values, with a few exceptions noted in the table comments.  When studies report asymmetric uncertainties (typically relevant only for the faintest few bins) we adopt the maximum as our fiducial uncertainty, as the \textit{emcee} procedure discussed below requires symmetric uncertainties.

\begin{deluxetable*}{ccccc}
\tabletypesize{\small}
\tablecaption{Literature Data Used to Constrain Luminosity Function Evolution at $z=$7--9}
\tablewidth{\textwidth}
\tablehead{
\colhead{Redshift} & \colhead{Reference} & \colhead{M$_{UV}$ Range} & {Primary Survey$^{\dagger}$} & \colhead{Type}}
\startdata
7.0&\citet{bowler14}&[$-$22.7,$-$21.8]&UltraVISTA/UKIDSS&Star-Forming Galaxies\\
7.0&\citet{atek15}&[$-$20.3,$-$15.3]&Frontier Fields&Star-Forming Galaxies\\
7.0&\citet{finkelstein15}&[$-$22.0,$-$18.0]&CANDELS&Star-Forming Galaxies\\
7.0&\citet{bouwens21}&[$-$22.2,$-$17.0]&CANDELS&Star-Forming Galaxies\\
7.0&\citet{bouwens22}&[$-$18.8,$-$13.8]&Frontier Fields&Star-Forming Galaxies\\
\hline
8.0&\citet{schmidt14}$^{a}$&[$-$22.0,$-$20.0]&BoRG&Star-Forming Galaxies\\
8.0&\citet{finkelstein15}&[$-$21.5,$-$18.5]&CANDELS&Star-Forming Galaxies\\
8.0&\citet{stefanon19}&[$-$22.5,$-$22.0]&UltraVISTA&Star-Forming Galaxies\\
8.0&\citet{bowler20}&[$-$22.9,$-$22.7]&UltraVISTA/VIDEO&Star-Forming Galaxies\\
8.0&\citet{bouwens21}&[$-$21.9,$-$17.6]&CANDELS&Star-Forming Galaxies\\
8.0&\citet{bouwens22}&[$-$18.8,$-$13.3]&Frontier Fields&Star-Forming Galaxies\\
\hline
9.0&\citet{mcleod16}&[$-$20.2,$-$17.5]&Frontier Fields&Star-Forming Galaxies\\
9.0&\citet{morishita18}&[$-$22.0,$-$21.0]&BoRG&Star-Forming Galaxies\\
9.0&\citet{stefanon19}&[$-$22.4,$-$21.6]&UltraVISTA/VIDEO&Star-Forming Galaxies\\
9.0&\citet{rojasruiz20}&[$-$22.0,$-$21.0]&BoRG&Star-Forming Galaxies\\
9.0&\citet{bowler20}&[$-$22.9,$-$21.9]&UltraVISTA/VIDEO&Star-Forming Galaxies\\
9.0&\citet{bouwens21}&[$-$21.9,$-$17.9]&CANDELS&Star-Forming Galaxies\\
9.0&\citet{bagley22}$^{a}$&[$-$23.0,$-$21.0]&BoRG/WISP&Star-Forming Galaxies\\
9.0&\citet{bouwens22}&[$-$18.8,$-$15.8]&Frontier Fields&Star-Forming Galaxies\\
9.0&\citet{finkelstein22}$^{a}$&[$-$22.5,$-$21.0]&CANDELS&Star-Forming Galaxies
\enddata
\tablecomments{Same as Table 1, for $z =$ 7--9.  $^{a}$Schmidt et al.\ (2014) did not bin their data, but did provide number densities in 0.5 magnitude bins upon request.  Finkelstein et al.\ (2022) and Bagley et al.\ (2022) implemented a continuous pseudo-binning method; here we use values every $\Delta$M$=$0.5 to be consistent with other datasets used.
}
\label{tab:data2}
\end{deluxetable*}

\begin{figure}[!t]
\epsscale{1.15}
\plotone{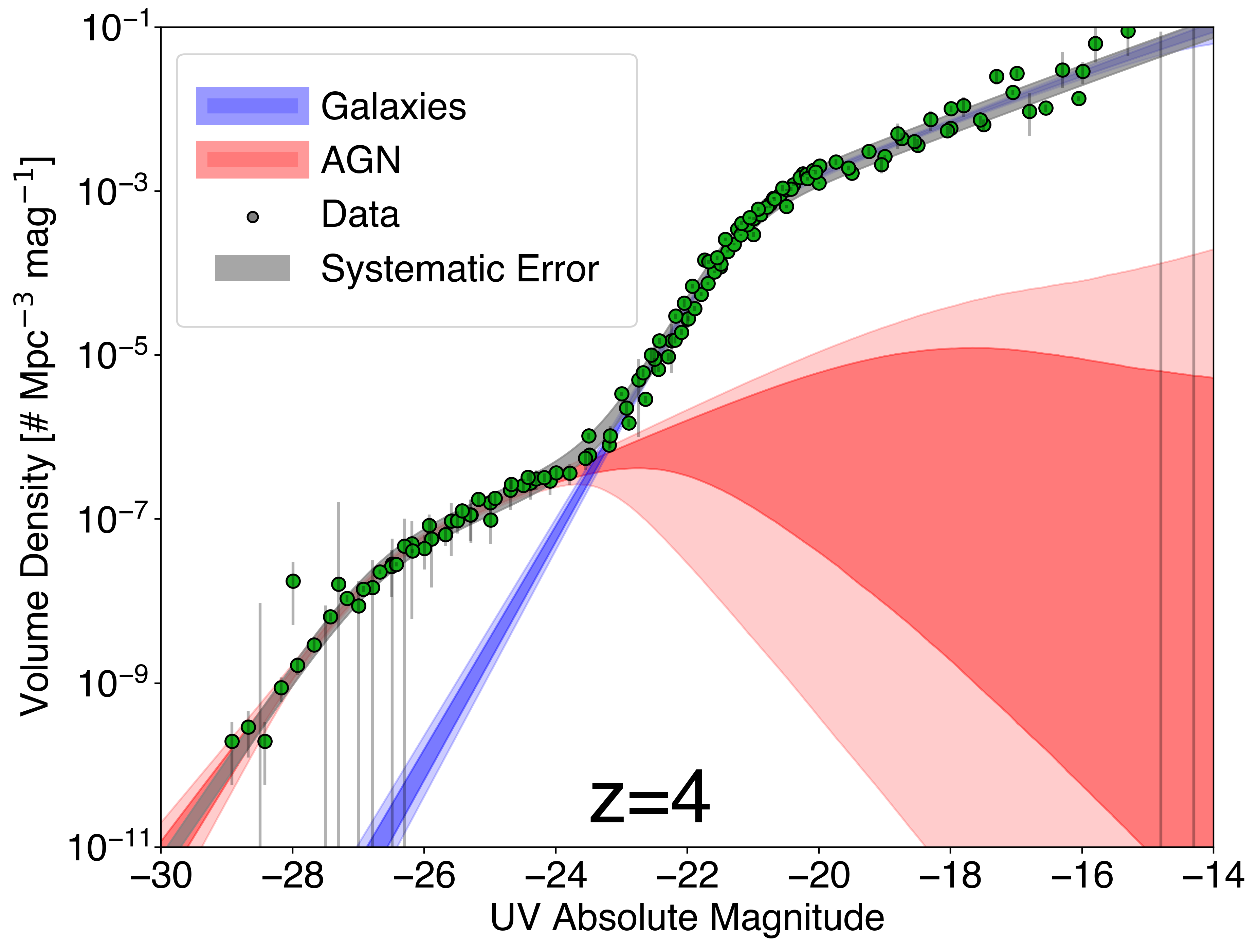}
\vspace*{-3mm}
\caption{The circles denote the observed UV luminosity function at $z =$ 4.  It is clear that the shape of these data are consistent with a functional form consisting of two double-power laws.  The shaded regions denote our 68\% (dark) and 95\% (light) confidence levels of our parameterized fit to these data (described in \S 2.3), which consists of separate double power-laws describing the AGN population (in red) and the star-forming galaxy population (in blue).  We include faint-end turnovers in these fits, though these are not yet well-constrained by the available data.  We also include a systematic error term which is proportional to the model value.  The median of this error term for the combined fit is shown by the gray shaded region.}
\label{fig:z4lf}
\end{figure}

\subsection{Form of the Luminosity Function}

As shown in Figure~\ref{fig:fig1}, the UV luminosity function exhibits multiple features.  We illustrate this in more detail in Figure~\ref{fig:z4lf}, highlighting $z =$ 4 where the data considered here are the most abundant.  It is apparent that the entire population of UV-luminous objects can be described by the sum of two double-power-law (DPL) functions, one each for AGN-dominated systems and star-formation-dominated systems.  At $M_{UV} > -$22, one can see the well-known galaxy UV luminosity function, described by a faint-end power-law slope, with a transition to a steeper decline at a characteristic magnitude.  Moving to brighter luminosities, the data transition again to a shallower slope, representing the faint-end of the AGN luminosity function,  transitioning again to a steeper slope brightward of a second characteristic magnitude.  While DPLs have been commonly used to describe the AGN population, the galaxy population has historically been fit as a Schechter function.  However, as discussed in \S 1 more recent results find that the bright end is modestly better fit by a shallower power-law decline rather than a steeper exponential decline.

We can quantify this shape by fitting a combination of two DPL functions to the data, one each for the AGN and star-forming galaxy population.  We fit the data for each population with this modified DPL function:
\begin{equation}
    \begin{split}
\phi(M)~=~&\phi^{\ast}~[10^{0.4 (\alpha + 1) (M - M^{\ast})}+10^{0.4 (\beta + 1) (M - M^{\ast})}]^{-1}\\
& \times [1 + 10^{0.4(M-M_t)\delta}]^{-1}
\end{split}
\end{equation}
\noindent where $\phi^{\ast}$ is the characteristic number density, $M^{\ast}$ is the characteristic magnitude, $\beta$ is the bright-end slope, and $\alpha$ is the faint-end slope\footnote{Some AGN studies use $\beta$ to refer to the faint-end slope, but we use $\alpha$ to be consistent with the standard for galaxy luminosity functions.}.  We include the possibility for a deviation of the faint-end slope at a magnitude $M_t$. The slope at $M_{UV} > M_t$ is modified by the parameter $\delta$, where $\delta =$ 0 would result in no deviation, and progressively more positive values would lead to a more dramatic shallowing (see also \citealt{jaacks13} and \citealt{bouwens17} for other ways to parameterize a turnover).  The full UV luminosity function is thus described by
\begin{equation}
    \phi(M) = \phi_{AGN}(M) + \phi_{galaxy}(M),
\end{equation}
\noindent requiring 12 free parameters to describe the data across the full dynamic range in observed UV luminosity. 

While Figure 2 shows that the data at $z =$ 4 span a wide enough dynamic range in luminosity ($\sim$15 magnitudes) to allow decent constraints on most parameters, the same is not true at $z \geq$ 7, where observations lack constraints at both the brightest and faintest regimes (Figure 1).  We therefore leverage previous evidence that both the galaxy and AGN luminosity function parameters evolve smoothly with redshift \citep[e.g.][]{finkelstein16,kulkarni19}, and fit the luminosity functions to all of $z =$ 3--9 simultaneously.  

We assume that all 12 luminosity function parameters can be described by a smoothly evolving polynomial as a function of (1$+z$).  We aim to construct the simplest model to describe the data, thus we assume in most cases that these functions are linear.  The exceptions are for $\phi_{AGN}^{\ast}$ and $M_{AGN}^{\ast}$ which we fit as quadratic functions, following \citet{kulkarni19}\footnote{\citet{kulkarni19} use a cubic function to describe $M_{AGN}^{\ast}$ over 0 $< z <$ 7, but at $z >$ 3, which we consider here, their results are consistent with a quadratic form.}.  This has the effect of allowing the bright-end of the AGN luminosity function to fall off rapidly with increasing redshift, consistent with the rapid drop in the number density of luminous quasars \citep[e.g.,][]{hopkins07}.

We note that our inclusion of a faint-end deviation from the power-law allows our functional form to be representative of the expected shape of the full luminosity function.  However, in practice the present data have little constraining power on these quantities.  On the galaxy side, observations do constrain any turnover to be at $M \gtrsim -$16 \citep[e.g.][]{livermore17,atek18,bouwens22}, though the uncertainties on the faintest bins are high leaving these studies unable to more precisely pin down any deviation.  On the AGN side, at $M > -$24 observations are dominated by star-forming galaxies, thus also cannot constrain in detail where any turnover in the AGN component might begin.  

Nevertheless, we include these parameters such that our constraints on the full luminosity function are inclusive of the uncertainties over where these turnovers occur.  While the galaxy turnover will have little effect on our results, the turnover in the AGN luminosity function could matter.  Either arbitrarily truncating the AGN luminosity function or allowing it to continue to very faint luminosities would artificially attribute a greater or lower respective fraction of the observed number densities to star forming galaxies.  Given the lack of observational evidence for the presence and evolution of these turnovers, we also elect to fit the redshift evolution of the turnover magnitude as a quadratic function of (1$+z$).  We note that future {\it JWST} observations can better constrain these turnovers (spectroscopic followup to identify faint AGNs, and deep imaging to push further down the star-forming galaxy luminosity functions).


\subsection{Constraining the Evolution of the Luminosity Function with MCMC}

Our model of the redshift evolution of the full UV luminosity function has 29 free parameters in total.  The evolution of the AGN luminosity function is described by the following terms:
\begin{equation}
\begin{split}
    log~\phi^{\ast}_a(z) = & \mathcal{P}_{a,0} + \mathcal{P}_{a,1}(1+z) + \mathcal{P}_{a,2}(1+z)^2\\
    M^{\ast}_a(z) = & \mathcal{M}_{a,0} + \mathcal{M}_{a,1}(1+z) + \mathcal{M}_{a,2}(1+z)^2\\
    \alpha_a(z) = & \mathcal{A}_{a,0} + \mathcal{A}_{a,1}(1+z)\\
    \beta_a(z) = & \mathcal{B}_{a,0} + \mathcal{B}_{a,1}(1+z)\\
    M_{t,a}(z) = & \mathcal{T}_{a,0} + \mathcal{T}_{a,1}(1+z) + \mathcal{T}_{a,2}(1+z)^2\\
    \delta_a(z) = & \mathcal{D}_{a,0} + \mathcal{D}_{a,1}(1+z)
\end{split}
\end{equation}
\noindent where the coefficients represent the 15 free parameters.  The numerical subscripts denote the constant (0), linear (1) and quadratic (2) terms, with the ``a" denoting that these represent the AGN luminosity function.  Likewise, the galaxy luminosity function can be described by these 13 coefficients, where the ``g" denotes that these represent the galaxy luminosity function:
\begin{equation}
\begin{split}
    log~\phi^{\ast}_g(z) = & \mathcal{P}_{g,0} + \mathcal{P}_{g,1}(1+z)\\
    M^{\ast}_g(z) = & \mathcal{M}_{g,0} + \mathcal{M}_{g,1}(1+z)\\
    \alpha_g(z) = & \mathcal{A}_{g,0} + \mathcal{A}_{g,1}(1+z)\\
    \beta_g(z) = & \mathcal{B}_{g,0} + \mathcal{B}_{g,1}(1+z)\\
    M_{t,g}(z) = & \mathcal{T}_{g,0} + \mathcal{T}_{g,1}(1+z) + \mathcal{T}_{g,2}(1+z)^2\\
    \delta_g(z) = & \mathcal{D}_{g,0} + \mathcal{D}_{g,1}(1+z)
\end{split}
\end{equation}

\begin{deluxetable}{cccc}
\vspace{2mm}
\tabletypesize{\small}
\tablecaption{Posterior Constraints on Free Parameters}
\tablewidth{\textwidth}
\tablehead{
\colhead{Parameter} & \colhead{Median} & \colhead{16 percentile} & \colhead{84 percentile}}
\startdata
$\mathcal{P}_{a,0}$ & $-$6.94 & $-$8.37 & $-$5.46 \\
$\mathcal{P}_{a,1}$ & 0.65 & 0.11 & 1.15 \\
$\mathcal{P}_{a,2}$ & $-$0.15 & $-$0.19 & $-$0.1 \\
$\mathcal{M}_{a,0}$ & $-$26.65 & $-$29.21 & $-$23.84 \\
$\mathcal{M}_{a,1}$ & 0.36 & $-$0.6 & 1.24 \\
$\mathcal{M}_{a,2}$ & $-$0.08 & $-$0.15 & 0.01 \\
$\mathcal{A}_{a,0}$ & $-$1.45 & $-$2.02 & $-$0.82 \\
$\mathcal{A}_{a,1}$ & $-$0.08 & $-$0.19 & 0.03 \\
$\mathcal{B}_{a,0}$ & $-$2.14 & $-$4.21 & $-$0.61 \\
$\mathcal{B}_{a,1}$ & $-$0.37 & $-$0.66 & 0.04 \\
$\mathcal{P}_{g,0}$ & $-$1.45 & $-$1.68 & $-$1.27 \\
$\mathcal{P}_{g,1}$ & $-$0.31 & $-$0.35 & $-$0.27 \\
$\mathcal{M}_{g,0}$ & $-$21.18 & $-$21.42 & $-$20.97 \\
$\mathcal{M}_{g,1}$ & 0.02 & $-$0.02 & 0.06 \\
$\mathcal{A}_{g,0}$ & $-$1.27 & $-$1.36 & $-$1.19 \\
$\mathcal{A}_{g,1}$ & $-$0.11 & $-$0.12 & $-$0.09 \\
$\mathcal{B}_{g,0}$ & $-$4.79 & $-$5.13 & $-$4.45 \\
$\mathcal{B}_{g,1}$ & 0.05 & $-$0.0 & 0.1 \\
$\mathcal{T}_{a,0}$ & $-$18.47 & $-$31.56 & $-$7.15 \\
$\mathcal{T}_{a,1}$ & $-$0.83 & $-$4.36 & 3.7 \\
$\mathcal{T}_{a,2}$ & 0.07 & $-$0.26 & 0.33 \\
$\mathcal{D}_{a,0}$ & 2.09 & $-$0.12 & 4.34 \\
$\mathcal{D}_{a,1}$ & $-$0.01 & $-$0.3 & 0.27 \\
$\mathcal{T}_{g,0}$ & $-$21.32 & $-$29.47 & $-$14.91 \\
$\mathcal{T}_{g,1}$ & 2.26 & 0.15 & 5.1 \\
$\mathcal{T}_{g,2}$ & $-$0.16 & $-$0.37 & $-$0.01 \\
$\mathcal{D}_{g,0}$ & $-$0.76 & $-$1.05 & $-$0.51 \\
$\mathcal{D}_{g,1}$ & 0.19 & 0.13 & 0.27 \\
ln $\mathcal{F}$ & $-$1.37 & $-$1.43 & $-$1.32 \\
\enddata
\tablecomments{The median and 68\% confidence range on our 29 free parameters from our \textsc{emcee} fit to the observed data over $z =$ 3--9.}
\label{tab:zparams}
\end{deluxetable}

Together, the combination of luminosity functions can be described by these 28 free parameters.  We include one additional parameter, a systematic error term $\mathcal{F}$, which is a single multiplicative factor of the combined luminosity function model.  This allows flexibility in our model fitting should the data have underestimated uncertainties.  There is evidence this may be the case, as the scatter in data points between different studies at the same magnitude often exceeds the uncertainties published in these studies.  This can be seen in the faint-end at $z =$ 4 in Figure 2.

\begin{deluxetable*}{ccccccccc}
\vspace{2mm}
\tabletypesize{\small}
\tablecaption{Posterior Constraints on Derived Luminosity Function Parameters}
\tablewidth{\textwidth}
\tablehead{
\colhead{Parameter} & \colhead{Prior} & \colhead{z=3} & \colhead{z=4} & \colhead{z=5} & \colhead{z=6} & \colhead{z=7} & \colhead{z=8} & \colhead{z=9}}
\startdata
log $\phi^{\ast}_{AGN}$ &[$-$20, $-$4]&$ -6.72_{-0.27}^{+0.31} $&$ -7.38_{-0.21}^{+0.23} $&$ -8.37_{-0.2}^{+0.25} $&$ -9.64_{-0.3}^{+0.37} $&$ -11.18_{-0.52}^{+0.59} $&$ -13.02_{-0.85}^{+0.92} $&$ -15.16_{-1.25}^{+1.35} $\\
$M^{\ast}_{AGN} $&[$-$34, $-$22]&$ -26.45_{-0.39}^{+0.53} $&$ -26.75_{-0.29}^{+0.34} $&$ -27.26_{-0.25}^{+0.35} $&$ -27.91_{-0.38}^{+0.5} $&$ -28.71_{-0.71}^{+0.85} $&$ -29.66_{-1.2}^{+1.38} $&$ -30.75_{-1.86}^{+2.09} $\\
$\alpha_{AGN} $&[$-$4, 0]& $ -1.76_{-0.39}^{+0.21} $&$ -1.83_{-0.29}^{+0.15} $&$ -1.91_{-0.25}^{+0.16} $&$ -2.0_{-0.38}^{+0.24} $&$ -2.09_{-0.71}^{+0.34} $&$ -2.17_{-1.2}^{+0.44} $&$ -2.25_{-1.86}^{+0.54} $\\
$\beta_{AGN} $&[$-$8, $-$2]&$ -3.61_{-0.39}^{+0.45} $&$ -3.94_{-0.29}^{+0.31} $&$ -4.32_{-0.25}^{+0.48} $&$ -4.7_{-0.38}^{+0.85} $&$ -5.08_{-0.71}^{+1.25} $&$ -5.46_{-1.2}^{+1.67} $&$ -5.84_{-1.86}^{+2.07} $\\
log $\phi^{\ast}_{gal}$&[$-$8, $-$1]&$ -2.72_{-0.08}^{+0.07} $&$ -3.03_{-0.07}^{+0.06} $&$ -3.35_{-0.06}^{+0.06} $&$ -3.66_{-0.08}^{+0.08} $&$ -3.97_{-0.11}^{+0.12} $&$ -4.29_{-0.14}^{+0.17} $&$ -4.6_{-0.17}^{+0.21} $\\
$M^{\ast}_{gal} $&[$-$24, $-$16]&$ -21.13_{-0.09}^{+0.08} $&$ -21.11_{-0.07}^{+0.07} $&$ -21.1_{-0.07}^{+0.07} $&$ -21.08_{-0.09}^{+0.09} $&$ -21.06_{-0.12}^{+0.13} $&$ -21.05_{-0.15}^{+0.16} $&$ -21.03_{-0.19}^{+0.2} $\\
$\alpha_{gal}$&[$-$4, 0]&$ -1.71_{-0.09}^{+0.03} $&$ -1.82_{-0.07}^{+0.03} $&$ -1.93_{-0.07}^{+0.03} $&$ -2.04_{-0.09}^{+0.04} $&$ -2.15_{-0.12}^{+0.05} $&$ -2.26_{-0.15}^{+0.06} $&$ -2.37_{-0.19}^{+0.08} $\\
$\beta_{gal} $&[$-$8, $-$2]&$ -4.59_{-0.09}^{+0.15} $&$ -4.54_{-0.07}^{+0.11} $&$ -4.49_{-0.07}^{+0.09} $&$ -4.44_{-0.09}^{+0.1} $&$ -4.39_{-0.12}^{+0.13} $&$ -4.33_{-0.15}^{+0.17} $&$ -4.28_{-0.19}^{+0.21} $\\
\hline
$M_{t,AGN} $&[$-24$, $-$16]&$ -20.81_{-0.39}^{+3.26} $&$ -20.41_{-0.29}^{+2.14} $&$ -20.65_{-0.25}^{+2.75} $&$ -20.74_{-0.38}^{+3.04} $&$ -20.58_{-0.71}^{+2.63} $&$ -20.19_{-1.2}^{+1.87} $&$ -19.87_{-1.86}^{+2.6} $\\
$\delta_{AGN} $&[0, 4]&$ 2.03_{-0.39}^{+1.35} $&$ 2.01_{-0.29}^{+1.13} $&$ 2.01_{-0.25}^{+0.92} $&$ 2.0_{-0.38}^{+0.86} $&$ 1.97_{-0.71}^{+0.96} $&$ 1.94_{-1.2}^{+1.17} $&$ 1.91_{-1.86}^{+1.41} $\\
$M_{t,gal} $&[$-$16, $-$10]&$ -15.09_{-0.09}^{+1.5} $&$ -14.0_{-0.07}^{+1.4} $&$ -13.48_{-0.07}^{+1.82} $&$ -13.33_{-0.09}^{+1.97} $&$ -13.45_{-0.12}^{+1.72} $&$ -13.98_{-0.15}^{+1.42} $&$ -15.4_{-0.19}^{+2.45} $\\
$\delta_{gal} $&[0, 4]&$ 0.01_{-0.09}^{+0.03} $&$ 0.23_{-0.07}^{+0.1} $&$ 0.43_{-0.07}^{+0.17} $&$ 0.62_{-0.09}^{+0.22} $&$ 0.8_{-0.12}^{+0.29} $&$ 0.99_{-0.15}^{+0.36} $&$ 1.18_{-0.19}^{+0.43} $\\
\enddata
\tablecomments{The median and 68\% confidence range of our fiducial model evaluated at seven integer redshifts.  The horizontal line distinguishes the parameters which describe the luminosity function turnovers, which are not well-constrained given the observations (though our assumption of smooth evolution of these quantities with redshift does restrict the uncertainties from being too large).}
\label{tab:lfparams}
\end{deluxetable*}

We constrain the posterior distribution of this complete set of 29 free parameters using the \textsc{emcee} Markov Chain Monte Carlo (MCMC) python package \citep{foreman-mackey13}.  We assume that the uncertainties on the observed number densities are Gaussian, and that we can describe the likelihood of a given model via:
\begin{equation}
    \begin{split}
    & \mathrm{ln} (p | D(z), \sigma, \Theta) = \\
    & -0.5 \sum_{z=3}^{z=9} \sum_n \left[\frac{D(z)_n - \phi(\Theta)}{\sigma_{n,tot}^2} + ln(2 \pi \sigma_{n,tot}^2)\right] P(\phi[\Theta])
\end{split}
\end{equation}
\noindent where $p$ represents the likelihood that a given set of data ($D[z]$) and associated uncertainties ($\sigma$) are represented by a model with a set of parameters $\Theta$.   To constrain the luminosity function data over all redshifts simultaneously, the likelihood function first sums over all data at a given redshift denoted by the subscript $n$, and then sums over all redshifts 3 $< z <$ 9.

We adopt a flat, uninformative prior $P(\phi[\Theta])$, where the prior is imposed on the 12 luminosity function terms, which the likelihood function calculates during each step given the set of free parameters for that step.  The term $\sigma_{n,tot}$ represents the total uncertainty, which is a combination of the uncertainty on the observed data and our systematic error term $\mathcal{F}$ (multiplied by the model):
\begin{equation}
\sigma_{n,tot}^2 = \sigma_n^2 + \phi(\Theta)^2 \mathcal{F}^2
\end{equation}
\noindent 

The \textsc{emcee} algorithm requires starting positions for each parameter to initialize the MCMC chains, and the algorithm will be the most efficient if these starting parameters are near to the likely true values.  To estimate these starting parameters we first use \textsc{emcee} to derive estimates of the 12 luminosity function parameters from Equations 1 and 2 at each redshift independently.  For these ``pre-processing'' \textsc{emcee} runs, we use 100 walkers and 100,000 steps, drawing a posterior from the last 10\% of the flattened chains.  We then fit linear (or quadratic where needed) functions to the median results as a function of (1$+z$) using \texttt{scipy} \texttt{minimize}.  The result of this pre-processing is an initial estimate of the 28 free parameters which describe the evolving luminosity function (we initialize the systematic error term $\mathcal{F}$ at a value of ln $\mathcal{F}=$0).

Using these starting positions, we fit the evolving luminosity function with our set of free parameters using 500 walkers and 500,000 steps.  These quantities were derived iteratively to ensure both reasonable acceptance fractions and that we could run the chain for large multiples of the auto-correlation times.  Our final chain had an acceptance fraction of 24.6\% and a median auto-correlation time of 5500 steps (for a ratio of the number of steps to the auto-correlation time of $\sim$90).  We derive a final posterior sample for each free parameter by discarding the first 3$\times$ the mean auto-correlation time, and thinning the remaining chain by half of this mean auto-correlation time.  We note that this mean auto-correlation time is 12500 steps.  This is significantly larger than the median as it is skewed by large auto-correlation times of $\sim$50,000 steps for the four terms which describe the turnover magnitude for both the AGN and star-forming galaxy components.  This is not surprising as these turnovers are not constrained by the data; the full chain is still $\sim$10$\times$ these few large auto-correlation times.  Excluding these four terms, the mean auto-correlation time is 6200 steps, close to the median for the full set of terms.

This final posterior consists of 30,600 chain steps for each of the 29 free parameters. Our results are summarized in two tables. Table 3 lists the posterior values of our free parameters, and Table 4 shows the resulting constraints on the luminosity function parameters at each redshift considered here, as well as the assumed prior values on these quantities.  

\section{Results}

\begin{figure*}[!t]
\epsscale{1.2}
\plotone{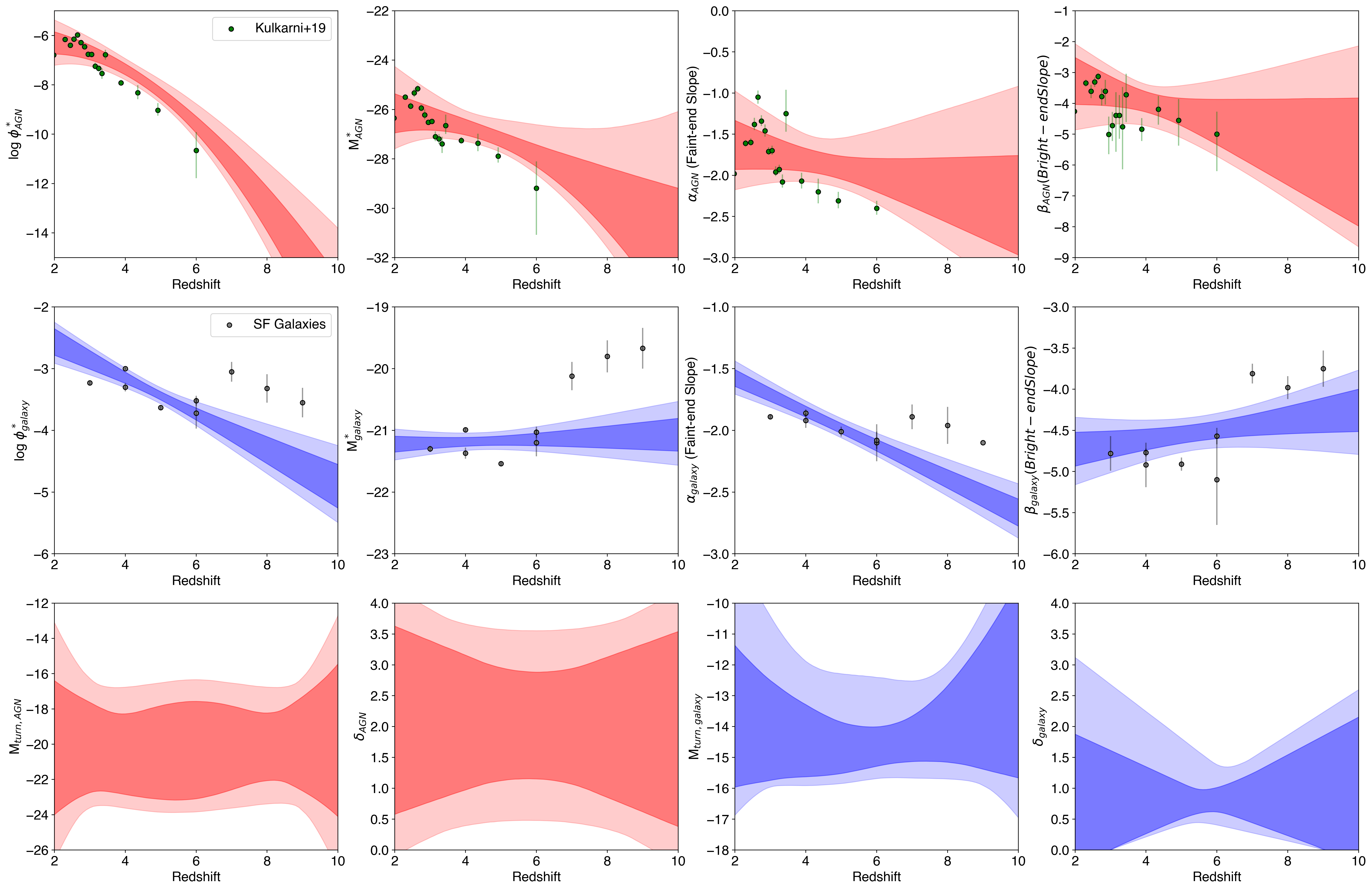}
\vspace*{-5mm}
\caption{The posterior constraints on the 12 luminosity function parameters as a function of redshift.  The dark and light shaded regions denote the 68 and 95\% confidence intervals, respectively.  Red colors denote AGN-dominated object parameters, while blue denotes star-forming galaxy parameters.  The data points shown in the top row are the direct fits to the AGN luminosity functions at individual redshifts from \citet{kulkarni19}.  We see excellent agreement at the lowest redshifts probed.  At $z \gtrsim$ 3, our fits prefer modestly higher $\phi^{\ast}$, fainter M$^{\ast}$ and steeper $\alpha$. This differences could be due to our methodology's ability to fit the AGN luminosity function without separating individual objects into AGN or star-forming galaxies, though the differences are only significant at the 1--2$\sigma$ level.  In the middle row, the data points shown come from \citet{harikane21} at $z =$ 3--7, \citet{adams20} at $z =$ 4, \citet{bowler15} at $z =$ 6, and \citet{bowler20} at $z =$ 8--9.  We see generally good agreement with the previous star-forming galaxy only luminosity functions to the results from our combined model at $z \leq$ 6.  At higher redshifts, our fits prefer lower  $\phi^{\ast}$, brighter M$^{\ast}$, steeper $\alpha$ and steeper $\beta$.  However, as shown in Figure~\ref{fig:lfs}, our fits match the observed data quite well, thus these discrepancies with previous results highlight the degeneracies between different luminosity function parameters (see Figure~\ref{fig:lf_comparison}).  The bottom row highlights that the turnovers of neither luminosity function are well constrained presently, though lensing studies do rule out a turnover at $M_{UV} \lesssim -$16.}
\label{fig:lfparams}
\end{figure*}

\subsection{Evolution of the Luminosity Function}

Figure~\ref{fig:lfparams} plots the posterior constraints on the evolution of the 12 parameters needed to describe the UV luminosity function (which themselves are described by the 29 free parameters from our \textsc{emcee} fit).  To first order, one can see that the majority of the luminosity function parameters are fairly well constrained.  The exception is the bottom row, which shows the four parameters describing the faint-end turnovers in both the AGN and star-forming galaxy components.  However, this is not surprising given the lack of data constraining these turnovers.  On the AGN side, the turnover is constrained to be between M$_{UV} \sim$ $-$24 to $-$18.  The bright limit is constrained by the data directly probing the faint-end slope of the AGN luminosity function, while the faint end is constrained by the data probing the faint-end slope of the galaxy population (e.g., the combined luminosity function could exceed the observations if the AGN component did not turn over by M$_{UV} \sim$ $-$18).  The star-forming galaxy luminosity function turnover is constrained to occur faintward of M$_{UV} \sim$ $-$16.  The constraints on $M_{t,g}$ are apparently tighter than on $M_{t,a}$.  This is driven by studies of the Hubble Frontier Fields, which constrain the bright limit of any turnover.

\begin{figure*}[!t]
\epsscale{1.2}
\plotone{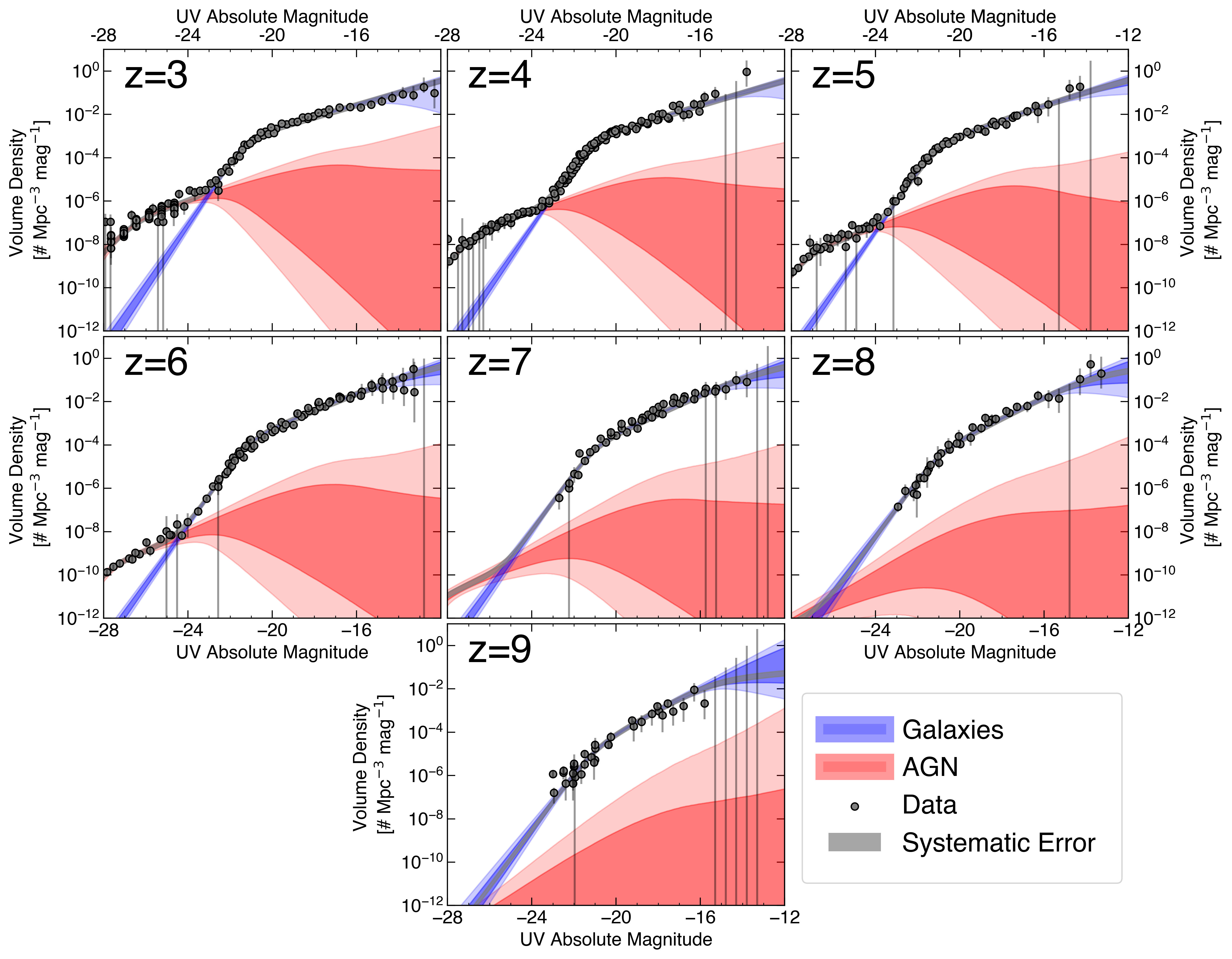}
\vspace*{-5mm}
\caption{The rest-frame UV luminosity functions at $z =$ 3--9.  The red (blue) color shaded regions show the posteriors on the AGN (star-forming galaxy) component, with the dark and light regions representing the 68\% and 95\% confidence intervals.  The gray shaded region is the total luminosity function, where the width visualizes the systematic error term included in the fitting.  The circles show the literature data at each redshift used to constrain our model.  This figure highlights that our smooth-evolution parameterization is able to reproduce the data at all redshifts considered, across a wide dynamic range in UV luminosity, placing reasonable constraints on the likely AGN luminosity function at $z \geq$ 7.}
\label{fig:lfs}
\end{figure*}

The top row describes the four parameters which comprise the AGN component of the luminosity function.  The constraints on these four parameters are the tightest at $z \leq$ 6, where the observations directly probe the AGN component.  While the uncertainties widen at $z \geq$ 7, our method of simultaneously fitting the luminosity functions at all redshifts allows us to place constraints on the plausible AGN luminosity function at $z \geq$ 7 for the first time (given our assumption of smooth evolution of the chosen parameters).  We find that the data prefer, with increasing redshift, a decreasing $\phi^{\ast}$, brightening $M^{\ast}$, a steepening faint-end slope $\alpha$, and a steepening bright-end slope $\beta$.

\begin{figure*}[!t]
\epsscale{1.2}
\plotone{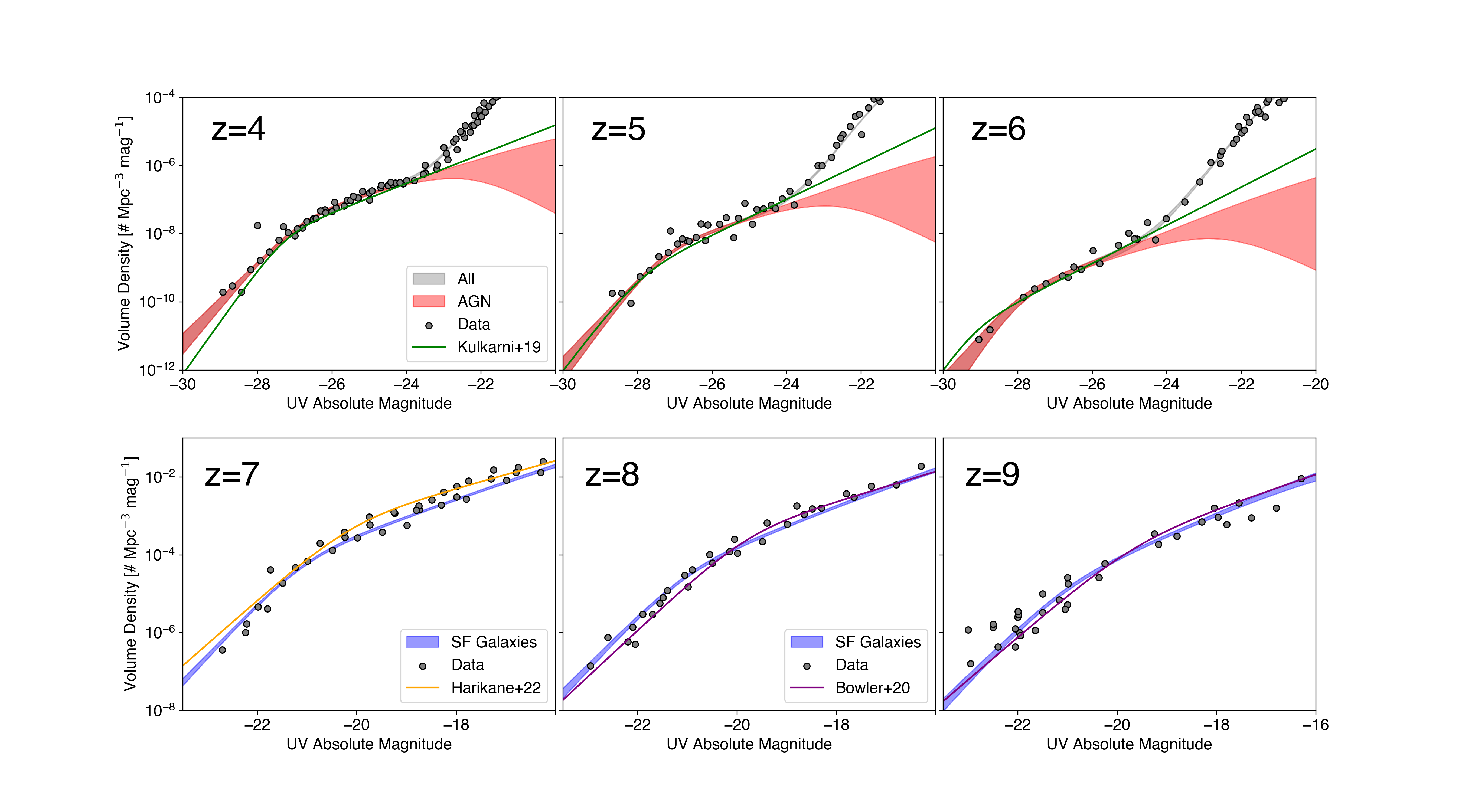}
\vspace*{-10mm}
\caption{Top: A comparison of our AGN luminosity function results (shaded region is the 68\% confidence interval) to the AGN luminosity functions from \citet{kulkarni19}.  While there are modest differences in the luminosity function parameters (Figure 3), the differences in $\phi^{\ast}$ and $M^{\ast}$ combine to have little effect on the luminosity function shape.  A more apparent difference is in the faint-end slope, where our method yields a flatter slope.  Bottom:  A comparison of our star-forming galaxy luminosity function results (shaded region is the 68\% confidence interval) to the galaxy luminosity functions from \citet{harikane21} and \citet{bowler20}.  While both previous works find a fainter value of $M^{\ast}$ , due to the degeneracies between these parameters both the published luminosity functions and those we derive here appear able to match the observations.}
\label{fig:lf_comparison}
\end{figure*}

The middle row describes the four parameters which make up the star-forming galaxy component of the luminosity function.  The constraints here are tighter than on the AGN parameters, reflecting the greater numbers of star-formation-dominated systems known across all redshifts.  Our model prefers, with increasing redshift, a decreasing $\phi^{\ast}$, relatively flat $M^{\ast}$, steepening $\alpha$, and a shallowing $\beta$.

Figure~\ref{fig:lfs} shows how these parameters combine to create the luminosity functions at each of the integer redshifts we consider here.  Our assumption of double-power law luminosity function parameters which smoothly vary with redshift is able to well-match the observations at all of $z =$ 3--9, and across all UV absolute magnitudes considered.  At $z =$ 3--6, this figure visualizes that the available data constrain both the bright and faint-ends of both luminosity function components.  At $z \geq$ 7, the available data still provide some constraints at magnitudes both brighter and fainter than the galaxy $M^{\ast}$.  While the data do not appear to constrain the AGN-dominated regime, our smooth-evolution parameterization of the AGN luminosity function still places reasonable constraints on the likely number densities of such objects.

\subsection{Comparison to Previous Results}

In the top four panels of Figure~\ref{fig:lfparams} we compare to results from \citet{kulkarni19}, who performed a similar analysis as we do here solely on the AGN component.  At $z <$ 4, we see fairly good agreement with this AGN-only study.  At $z >$ 4, our model prefers modestly higher $\phi^{\ast}$, fainter $M^{\ast}$ and a shallower faint-end slope ($\alpha$).  In the middle row of Figure~\ref{fig:lfparams}, we compare to star-forming galaxy luminosity function results, restricting our comparison to the wide-field studies of \citet{harikane21} for $z =$ 3--7, \citet{adams20} for $z =$ 4, \citet{bowler15} at $z =$ 6, and \citet{bowler20} at $z =$ 8--9, all which also fit double-power law forms of the luminosity function.  We find good agreement with these individual-redshift studies at $z <$ 6.  At higher redshifts, the results from these previous studies would suggest higher $\phi^{\ast}$, fainter $M^{\ast}$, and shallower faint-end and bright-end slopes.

\begin{figure*}[!t]
\epsscale{1.2}
\plotone{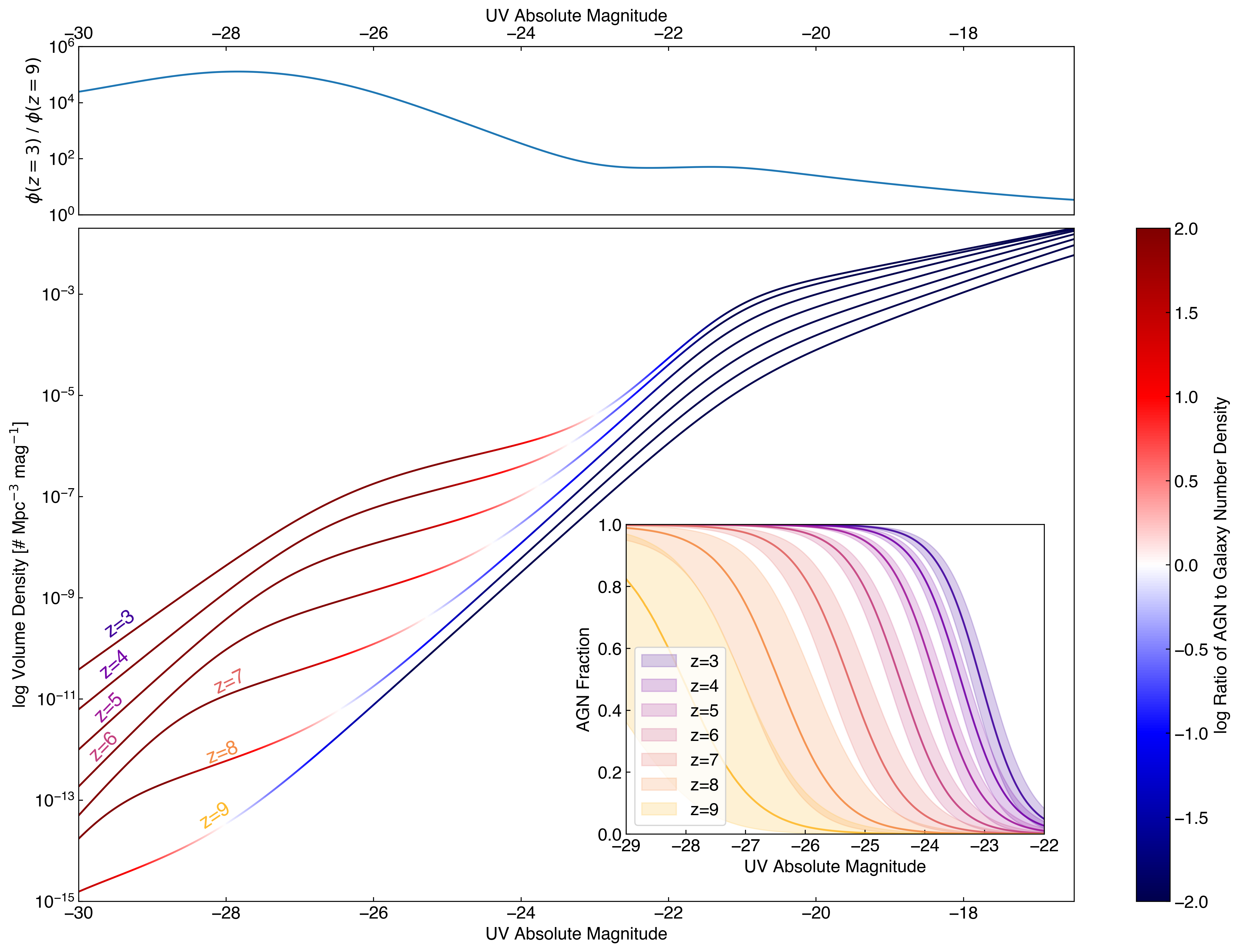}
\vspace*{-5mm}
\caption{The main panel shows the combined luminosity function at $z =$ 3--9 (redshift increases from top to bottom).  The line is shaded by the ratio of the AGN to galaxy number density at each absolute magnitude (color scale shown in the adjacent color bar).  The top panel shows the ratio of the $z =$ 3 to the $z =$ 9 luminosity function, highlighting the dramatic evolution at the AGN-dominated bright end, which increases by $\sim$five orders of magnitude at $M_{UV} = -$28 from $z =$ 9 to $z =$ 3, tracking the build-up of massive black holes in the early universe.  Conversely, the star-forming galaxy luminosity function exhibits a decline of only $\sim$two orders of magnitude at $M_{UV} = -$22.  The inset panel shows the AGN fraction (the ratio of the AGN number density to the total number density), highlighting the evolution of the transition point to AGN-dominated systems to brighter UV luminosities at higher redshifts.}
\label{fig:lfratio}
\end{figure*}

We visualize these results in Figure~\ref{fig:lf_comparison}.  The top row compares our results to the luminosity function fits from \citet{kulkarni19} at $z =$ 4--6.  The differences in $\phi^{\ast}$ and $M^{\ast}$ do not lead to significant differences in luminosity function shape between their study and our results.  We do however see differences in the faint-end slope, where the results from \citet{kulkarni19} lead to a higher abundance of faint AGN at $M_{UV} >-$ 24 than we find.  
However, their ability to obtain an accurate measurement of the abundance of faint AGNs is dependent on their estimate of the completeness of AGN selection at faint magnitudes, which is difficult due to the rising abundance of star-formation dominated systems at these luminosities.  As this is not required for our method, our results here imply a modestly lower abundance of faint-AGN than the \citet{kulkarni19} AGN-only study.

The bottom row of Figure~\ref{fig:lf_comparison} compares our luminosity function shapes to those of \citet{harikane21} at $z =$ 7 and \citet{bowler20} at $z =$ 8--9.  While these star-forming galaxy studies found a fainter $M^{\ast}$ (and correspondingly higher $\phi^{\ast}$) than we find, the differences in the luminosity function shape are fairly negligible.  This implies that differences in parameters between our results and these previous studies are not significant given the degeneracies between parameters.  Future deep wide-field surveys, such as COSMOS-Web with {\it JWST} (PI Casey) and the High Latitude Survey with {\it NGRST} will provide tighter constraints on the location and shape of the bright-end of the galaxy luminosity function at these high redshifts (see \S 4.5).

\section{Discussion}\label{sec:discussion}

\subsection{The Rise of AGNs in the Early Universe}
Our methodology allows us to draw inferences about the growth of both super-massive black holes and the rise of star formation across a wide range of cosmic time.  Figure~\ref{fig:lfratio} highlights our key findings.  The main panel of this figure shows the evolution of the full rest-UV luminosity function from $z =$ 3 to $z =$ 9 (showing the median at each redshift for clarity).  The lines are color-coded by the ratio of the AGN to the star-forming galaxy luminosity function at each magnitude.  These colors indicate that the luminosity corresponding to the transition from star-forming dominated systems to AGN dominated systems evolves with redshift, from $M_{UV} \sim -$23 at $z =$ 3 to $M_{UV} \sim -$28 at $z =$ 9.  At this highest redshift, observations of such bright objects have not yet been observed, indicating that all known $z \sim$ 9 systems are dominated by star formation (we discuss this more in \S 4.2).   The inset panel shows this result as the AGN fraction, calculated as the ratio of the AGN luminosity function to the total luminosity function at each redshift and absolute magnitude.  The location where this quantity is 0.5 is equivalent to the transition from red-to-blue in the main panel.

Combined, these figures highlight the rise of AGN-dominated systems in two ways.  First, the apparent dimming with decreasing redshift of the AGN-to-star-forming-galaxy transition point is a consequence of the significant rise in AGN activity in the universe.  At $z =$ 9, AGN-dominated objects exist in only the extreme brightest systems ($M_{UV} < -$26; not yet even confirmed to exist).  As redshift decreases, AGNs become more abundant with number densities overtaking star-formation-dominated systems at progressively fainter magnitudes.

The top panel of Figure~\ref{fig:lfratio} shows the ratio of the $z =$ 3 to $z =$ 9 UV luminosity functions.  As AGNs rise in prominence from essentially non-existent at $z =$ 9 to prominent at $z =$ 3, the UV luminosity function increases by $\sim$five orders of magnitude at $M_{UV} =$ -28.  Conversely, the well-studied star-forming galaxy UV luminosity function increases by only $\sim$2 orders of magnitude at $M_{UV} = -$22 across this same redshift range.  These different evolutionary trends are the consequence of two effects.  First, star-forming galaxies are relatively common at all redshifts, and the growth of stellar mass via star-formation is a roughly smooth process \citep{madau14}, meaning that individual star-forming galaxies grow brighter with time at a modestly slow pace \citep[e.g.][]{papovich11}.  When AGNs arrive on the scene, they are extremely bright.  Thus, even at low number densities, the onset of significant SMBH accretion results in significant changes to the shape of the extreme bright end of the rest-UV luminosity function.

\subsection{The Bright-end of the Observed $z =$ 9 Luminosity Function}
While the impending arrival of the first cycle of {\it JWST} data will unleash a flood of new knowledge about the $z =$ 9 universe, we already have our first glimpse of the UV luminosity function in this epoch.  While the faint-end appears to be continuing its march to modestly lower number densities and steeper faint-end slope \citep[e.g.,][]{mcleod16,oesch18,bouwens21}, with some controversy about whether the decline in number density from $z =$ 8 to 9 is accelerated compared to the decline from $z =$ 6 -- 8 (e.g., \citealt{oesch18} versus \citealt{mcleod16}), observations at the bright end paint a different picture.  The first hints that the abundance of bright $z =$ 9 galaxies were higher than expected came from a number of surveys based on {\it HST} pure parallel programs \citep[e.g.,][]{morishita18,rojasruiz20}, with additional evidence found from  wide-area ground-based studies \citep{bowler20}.  \citet{finkelstein22} placed constraints on the bright end of the $z =$ 9 UV luminosity function via the CANDELS survey, again finding a relatively high abundance, in agreement with the pure parallel studies.  Most recently, \citet{bagley22} performed an analysis on a combination of multiple pure parallel datasets, leveraging {\it Spitzer}/IRAC photometry to reduce contamination, finding again number densities of bright $z =$ 9 galaxies in agreement with the aforementioned studies (see also \citealt{robertsborsani22}).

\begin{figure}[!t]
\epsscale{1.22}
\plotone{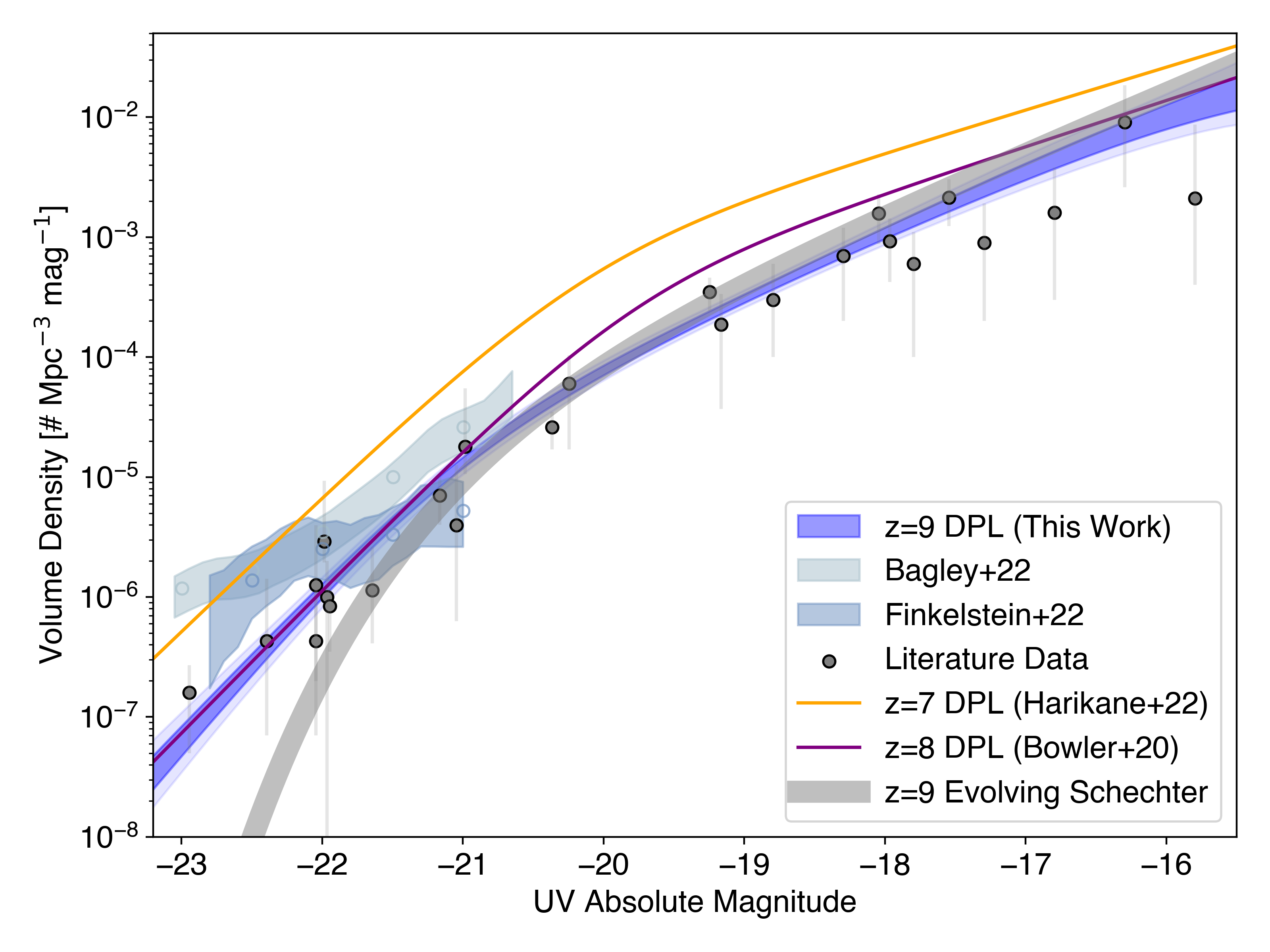}
\vspace*{-5mm}
\caption{A comparison of $z =$ 9 observations to the luminosity function from this work (shaded blue region).  Small symbols denote binned observations, while the light shaded regions denote the bright-end pseudo-binning results from \citet{finkelstein22} and \citet{bagley22} (faded open symbols denote the binned results from these two studies used to constrain our model).  The gray curve shows the evolving Schechter function of \citet{finkelstein16} at $z =$ 9. The observations and our results are consistent with this smoothly evolving function at $M > -$21, showing no evidence for an an accelerated decline in the UV luminosity function at $z >$ 8.  Results at the bright-end exceed this Schechter function, and are more consistent with our DPL form.  The number densities of bright ($M < -$21) galaxies appear to show little evolution from $z =$ 8 to 9.  This apparent excess of bright $z =$ 9 galaxies is unlikely to be due to AGN activity (Figure~\ref{fig:lfratio}) or a change in dust attenuation \citep{tacchella22}.  Spectroscopic confirmation of the bulk of these sources will come with {\it JWST}, and if they are validated, it could signal a change in the physics regulating star formation at the highest redshifts \citep[e.g.,][]{yung19a}, perhaps consistent with our inferred shallowing of the bright-end slope (Figure~\ref{fig:lfparams}).}
\label{fig:z9lf}
\end{figure}

Figure~\ref{fig:z9lf} shows the results from these studies, compared to both the predictions from the smoothly evolving Schechter function from \citet{finkelstein16}, and the measured double power-law luminosity functions at $z =$ 7 from \citet{harikane21} and $z =$ 8 from \citet{bowler20}.  As discussed above, the double-power law result from our work is a good match to the observed data, with nearly all observations consistent with the posterior of the model within 1$\sigma$.  The greatest difference is at the very bright end, $M < -$22, where both the \citet{finkelstein22} and \citet{bagley22} observations exceed the number densities from this work by $\sim$1--3$\sigma$.  Spectroscopic confirmation of these brightest $z =$ 9 sources are needed to verify these results.

Comparing our luminosity function posterior to previous fits at lower redshift, we recover an interesting trend, first noted by \citet{bowler20}.  While the evolution in the faint-end appears to be roughly smooth from $z =$7 $\rightarrow$ 8 and $z =$8 $\rightarrow$ 9, the same is not true at the bright end.  While there is significant evolution in the bright end from $z =$7 $\rightarrow$ 8, there appears to be little-to-no evolution in the bright end from $z =$ 8 $\rightarrow$ 9, as the \citet{bowler20} $z =$ 8 DPL goes right through our $z =$ 9 posterior at $M < -$21.

There are several potential explanations for this differential evolution.  Cosmic variance seems an unlikely explanation, as these results are based on large areas covering several different regions, including $>$100 {\it HST} pure parallel observations.  Another expectation could be differential evolution of dust attenuation.  While the faintest galaxies at $z \sim$ 6--8 already appear to be relatively unextincted, the brightest and most massive galaxies have shown evidence for significant dust attenuation \citep[e.g.][]{finkelstein12a,bouwens14}.  Should bright galaxies significantly grow their dust reservoirs from $z =$ 9 $\rightarrow$ 8, this could be a plausible explanation.  However, the recent results of \citet{tacchella22} have shown that bright $z =$ 9 galaxies exhibit similar rest-UV colors as bright $z =$ 4--8 galaxies, implying a similar level of dust attenuation.  

An exciting possibility would be if this slow bright-end evolution was due to a rapidly steepening AGN faint-end slope, such that galaxies at $M < -$22 were AGN-dominated.  However, as shown in Figure~\ref{fig:lfratio}, the results from this work implies that this is an unlikely explanation (though if the true luminosity function had a Schechter form, AGN would take over just brightward of existing observations; see \S 4.6).

It is also possible that the physics regulating star formation are evolving.  Recently, \citet{yung19a} found that a two-slope star-formation law in their semi-analytic model, which star-formation surface density and the gas surface density having an increasing slope (from 1 to 2) at high gas densities, is crucial to reproducing the bright-end at $z =$ 6--8. While such s star-formation law is motivated by observations of nearby star-forming regions, it is possible that the ``star-formation law'' continues to evolve to even higher redshifts, with a continued steepening (or increased normalization) resulting in an increased abundance of bright, very high-redshift galaxies.

The least exciting explanation would be if the majority of these sources were low-redshift contaminants.  A majority is unlikely as the studies used here were extremely careful in their sample selection, and several sources are spectroscopically confirmed \citep{zitrin15,oesch16,larson22}.  Nonetheless, this will remain a possibility until these galaxies are spectroscopically confirmed.  This should come in the near future, as many of these sources are targets of approved {\it JWST} Cycle 1 programs (PID 2426, PI Bagley and Rojas-Ruiz; PID 1345, 1758, PI Finkelstein; PID 1747, PI Roberts-Borsani).

Regardless of the explanation our results are fully consistent with a continued smooth decline in the UV luminosity functions at $z >$ 8.  The gray-shaded region shows the expected luminosity function at $z =$ 9 from the smoothly evolving Schechter function of \citet{finkelstein16}.  This curve is consistent with the data at all $M > -$ 21 (though the recent faint-end lensing results from \citealt{bouwens22} are systematically lower at $\sim$1$\sigma$ significance),  The only deviation is at the bright end, where the observations are higher and not lower as would be expected in an accelerated decline scenario.  These higher observations are easily explained by the luminosity function having a DPL form.

\begin{figure*}[!t]
\epsscale{1.2}
\plotone{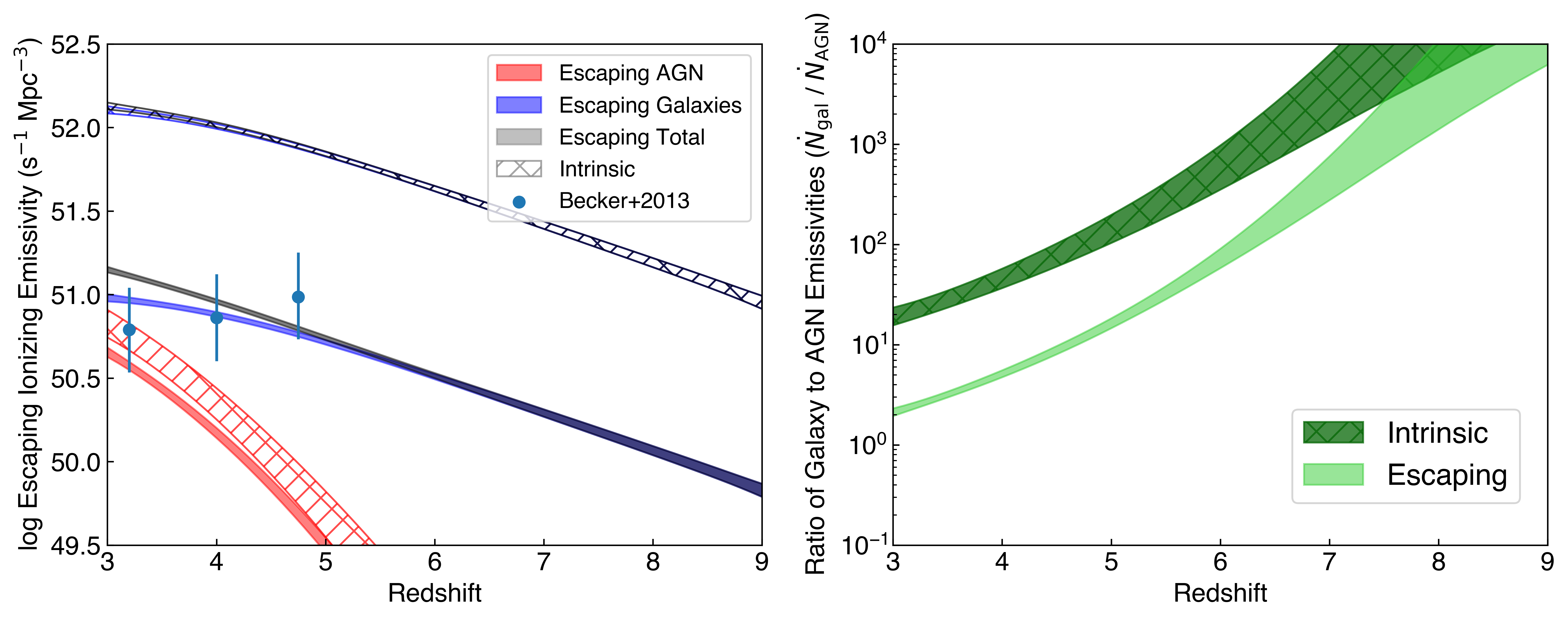}
\vspace*{-4mm}
\caption{Left) The evolution of the inferred ionizing emissivity versus redshift.  The open hatched regions show the intrinsically produced ionizing emissivity prior to any ISM/CGM/IGM gas absorption.  The solid shaded regions show the emissivities after we fold in our fiducial escape fractions (f$_{esc, galaxies} =$ 0.075, $f_{esc, AGN, M > -24} =$ 0.15, $f_{esc, AGN, M < -24} =$ 1.0).  These escape fraction values were chosen to roughly match the observed emissivities at $z =$ 3--5 \citep{becker13}.  Right) The ratio of emissivities from star-forming galaxies to that of AGNs.  Star-forming galaxies dominate the intrinsic emissivities at all redshifts considered, with an emissivity ratio of $\sim$20 at $z >$ 3, exceeding 10$^3$ in the epoch of reionization.  After application of our fiducial escape fractions, galaxies still dominate at all epochs, though only by a factor of $\sim$2 at $z =$ 3.}
\label{fig:emissivity}
\end{figure*}

Finally, we note that the observed $z =$ 9 luminosity function in the literature has a roughly single power-law appearance.  While our posterior DPL luminosity function has a noticeable break as shown in Figure~\ref{fig:z9lf}, the presently sizable observational uncertainties would easily hide this minor inflection in shape.  Our results show why it has been observationally difficult to pin down this break magnitude -- as one moves to $z =$ 9, the faint-end slope not only steepens, but the bright-end slope also shallows.  Thus, not only do the physical processes regulating the faint-end (e.g., stellar feedback) appear to be declining as has been previously noted, but feedback must also be decreasing at the bright end to enable a shallowing slope.  While it has previously been thought that AGN feedback, which is assumed to regulate the shape of the bright end, was not relevant at $z >$ 4 \citep[e.g.,][]{stevans18}, the changing shape of the bright end of the luminosity function out to $z =$ 9 points to a change in some physical process.  If AGN feedback were more prevalent at $z >$ 4 than previously thought, increasing AGN feedback from $z =$ 8 to 4 could explain the observations of massive galaxies shutting off their star-formation by $z \approx$ 4 \citep[e.g.,][]{glazebrook17,valentino20,stevans21}.

\subsection{Ionizing Emissivity}
The integral of the rest-UV luminosity function provides the total non-ionizing UV specific luminosity density $\rho_{UV}$ (units of erg s$^{-1}$ Hz$^{-1}$ Mpc$^{-3}$).  The ionizing emissivity can be estimated from this quantity given assumptions about the spectral shape of the population in question.  To explore the evolution of the \emph{intrinsic} (not accounting for ISM/CGM/IGM absorption) ionizing emissivity, we perform the following calculations.  As we explicitly allow turnovers in our luminosity function, we perform these limiting integrals to a value of $M_{UV} = -$10.  For star-forming galaxies, we assume an average conversion factor from $\rho_{UV}$ to ionizing emissivity of $\xi_{ion} =$ 25.6.  This value is motivated by increasing evidence that $\xi_{ion}$ increases with increasing redshift \citep[e.g.,][]{bouwens16b,stark17,endsley21,stefanon22}, and is consistent with the results of the empirical model of \citet{finkelstein19} for the redshift range considered here.   The intrinsic ionizing emissivity from galaxies is thus:
\begin{equation}
\dot{N}_{galaxies,intrinsic} = \rho_{UV,galaxies} \times \xi_{ion}
\end{equation}

We follow \citet{kulkarni19} to calculate the AGN ionizing emissivity, assuming that AGN exhibit a broken power law spectrum with a slope 0.61 redward of the Lyman break, and an AGN \ion{H}{1} ionizing spectral index of $\alpha_{QSO} =$ 1.7 \citep{lusso15}.  Combining Equations 6 and 7 from \citet{becker13}, we calculate the intrinsic AGN ionizing emissivity as:
\begin{equation}
\dot{N}_{AGN,intrinsic} = \rho_{UV,AGN} \left(\frac{912}{1500}\right)^{0.61} \frac{1}{h \alpha_{QSO}}
\end{equation}
where the first term is the integral of the luminosity function, the second term converts that specific luminosity density from the observed rest-wavelength to the Lyman limit, and the third term converts this quantity to photons (where $h$ is the Planck constant), and is an analytical solution to Equation 7 from \citet{becker13}.  Comparing Equation 8 to 7 shows that these assumptions lead to an effective value of $\xi_{ion, AGN} =$ 25.82.

We show these intrinsic ionizing emissivities as the open hatched regions in Figure~\ref{fig:emissivity}.  These emissivities rise with decreasing redshift for both AGNs and star-forming galaxies, mirroring the evolution of the luminosity function parameters.  In the right panel of Figure~\ref{fig:emissivity} we show the ratio of the galaxy to AGN emissivities, highlighting that galaxies significantly dominate the intrinsic emissivities at all redshifts considered here.  

\begin{figure*}[!t]
\epsscale{1.2}
\plotone{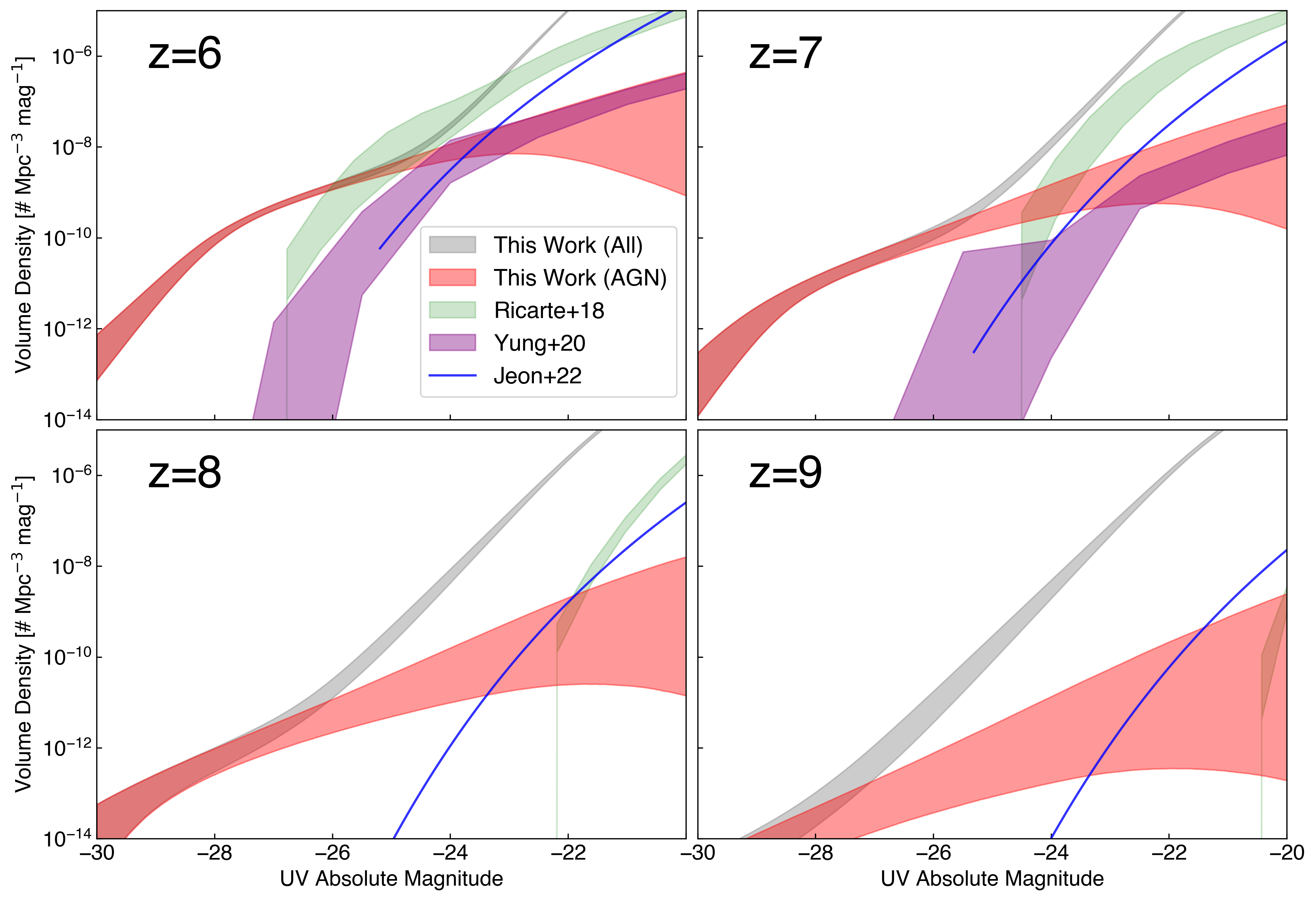}
\vspace*{-5mm}
\caption{A comparison of our derived AGN luminosity functions to three theoretical predictions from the literature, each with different assumptions about black hole seeding and accretion. All three models agree reasonablu well ($\sim$1--2 dex) with each other, as well as with observations at $M \sim -$24.  Some models over-predict the abundance of faint AGN, which could be rectified if much of this population was heavily obscured.  All three models are unable to predict the abundance of bright AGN in the early universe, which could be due to a combination of simulation volume limitations and the necessity of additional accretion mechanisms (e.g., super-Eddington accretion).}
\label{fig:bh}
\end{figure*}

While we leave a detailed investigation into potential ionizing photon escape fractions and the impact on reionization histories to future work, we do consider what simple escape fractions could be assumed which would replicate observed emissivities.  The solid shaded regions in both figures show the results if we assume that bright ($M < -$24) AGNs have a unity escape fraction, faint AGNs have an escape fraction of 15\%, and all galaxies have an escape fraction of 7.5\% (while galaxy escape fractions certainly may vary with luminosity, this is essentially an average escape fraction for the entire population).  These values were tuned such that the total emissivity was consistent with the observed emissivities from \citet{becker13}.  While the escaping ionizing emissivity from star-forming galaxies is significantly reduced, it still dominates over AGNs all the way to $z =$ 3 (albeit only a ratio of $\sim$2 at $z =$ 3, compared to $\sim$20 prior to escape).

These results contrast with those from the empirical model of \citet{finkelstein19}, who found that given their assumptions, the escaping ionizing emissivity from AGNs dominated over galaxies at $z <$ 4.6, with AGNs still contributing $\sim$10\% of the total emissivity at $z \sim$ 7.  However our methods differ as they did not consider constraints on the AGN luminosity function; rather, they considered a range of AGN emissivity evolution scenarios within the range of previously published results.  As we do not consider whether the assumptions made here are sufficient to satisfy constraints on the timeline of reionization, we cannot yet say which set of results are more likely.  An updated empirical reionization model including constraints on both the AGN and star-forming galaxy luminosity functions would lead to significant progress understanding the relative contribution of both populations.

\subsection{Constraints on Black Hole Seeding Models}
In Figure~\ref{fig:bh} we compare our AGN luminosity functions to the inferred rest-UV luminosity functions from three theoretical predictions with different underlying black hole seeding models.  \citet{ricarte18} created a semi-analytic model where they considered two seeding mechanisms (heavy and light).  They also considered two accretion scenarios -- a ``main sequence" scenario, where SMBH accretion tracks the star-formation rate, and a ``power law" scenario where they assume a power-law Eddington ratio distribution tuned to reproduce local observations.  At the redshifts we consider here which overlap with their work ($z =$ 6--9), there are not significant differences between the luminosity functions from these scenarios, so we show the light seed, power-law accretion scenario.  

\citet{yung21} also ran a semi-analytic model, seeding all top-level halos in their merger trees with 10$^4$ M\sol\ massive seeds, forward-modeling BH masses and accretion rates into synthetic SEDs, and measuring AGN UV luminosity functions at 2 $< z <$ 7.  These SMBHs grow rapidly during the radiatively efficient `bright mode' of AGN activity fueled by cold gas accretion from mergers and disk instabilities.  They also include a radiatively inefficient `radio' mode, though find that it is insignificant at high redshift.
Finally, \citet{jeon22} developed an analytic framework to study the maximal X-ray feedback from AGN allowable in the high-redshift universe.  They assume that every halo has a SMBH with a mass of a fixed fraction of the halo mass, showing that this fairly simple assumption produces luminosity functions in excellent agreement with published constraints on the high-redshift AGN luminosity functions.  While \citet{yung20b} published rest-UV luminosity functions, both \citet{ricarte18} and \citet{jeon22} publish bolometric luminosity functions.  We convert these to UV luminosity functions using the bolometric-to-UV corrections from \citet{shen20}.

Although these three models make a variety of seeding and SMBH growth assumptions, at redshifts where they overlap the agreement is within $\sim$1 dex.  Comparing to our results at $z =$ 6-7 where all three models provide predictions, we can see that the Yung et al.\ and Jeon et al.\ models have comparable number densities as the observations at $M \sim -$24, while the Ricarte et al.\ model is higher.  The \citet{yung21} SAM best matches our inferred AGN luminosity function's normalization and faint-end slope.  The \citet{ricarte18} model has a similar faint-end slope, albeit with a higher normalization.  The \citet{jeon22} model has both a steeper slope and higher normalization.  As our observations probe only the emitted UV light, an over-prediction of the models could be alleviated if a substantial fraction of the fainter AGNs were obscured by dust.  These dusty quasars would contribute to the bolometric luminosity function, but would not contribute to the escaping ionizing emissivity from AGNs, as the ionizing photons would likely also be absorbed by the dust.

We do see that all three models fail to produce the brightest AGNs, falling substantially below the observations at $M < -$26.  For the two SAMs, this is at least in part a volume limitation, as these are based on cosmological boxes with a fixed size.  Additionally, these models did not include super-Eddington accretion, which may be needed to create the most massive SMBHs in the early universe. The analytic \citet{jeon22} model has no such volume limitation, though their model's reliance on the Press-Schechter mass function results in an under-prediction of massive halos which would host these bright AGN.

\subsection{Predictions for Wide-Field Space Telescope Surveys}
Our methodology allows us to predict the abundance of rare, bright AGNs and star-forming galaxies, which (at the highest redshifts) are at volume densities not yet probed observationally.  The second half of this decade will see the advent of both ESA's {\it Euclid Observatory} and NASA's {\it Nancy Grace Roman Space Telescope}.  Both telescopes are aimed at wide-field near-infrared surveys.  When combined with onboard (in the case of {\it Euclid}) or ground-based Vera Rubin Observatory (VRO) Legacy Survey of Space and Time (LSST; in the case of \emph{Roman}) optical imaging, these surveys will allow significant numbers of rare bright high-redshift sources to be discovered.

To forecast the expected yield of such systems, we consider four surveys.  First, {\it Euclid} has two relevant planned surveys. The {\it Euclid} Wide survey will cover 14,500 deg$^2$ to an expected $YJH =$ 24.5 \citep[5$\sigma$, ][]{scaramella22}, while the {\it Euclid} Deep survey will cover 40 deg$^2$ to $YJH =$ 26.4 \citep[5$\sigma$, ][]{vanmierlo22}.  For \textit{NGRST} we consider here the planned 2277 deg$^2$ High Latitude Survey (HLS; 5$\sigma =$ 26.5) and a hypothetical 1 deg$^2$ ultra-deep survey to 5$\sigma =$ 29.5 \citep[e.g.,][]{drakos22,bagley22b}.  Estimation of robust photometric redshifts will require a fainter limiting magnitude in the dropout band to constrain the Lyman break.  As presently scoped these surveys will have roughly equal depth in all filters, thus we assume an effective depth 0.5 mag brighter than the stated depth for each survey.

We calculate the predicted number of objects by integrating the luminosity functions to the assumed effective depth and area for a given survey in redshift bins of $\Delta z =$ 0.1.  We then calculate the predicted number of objects above a given redshift by summing up these values above that given redshift, up to $z =$ 10.  To incorporate our uncertainties, we do this calculation for each output step in our \textsc{emcee} chain and calculate the 16/50/84 percentiles.  These calculations are done separately for the AGN and star-forming galaxy luminosity functions. 

\begin{figure*}[!t]
\epsscale{0.55}
\plotone{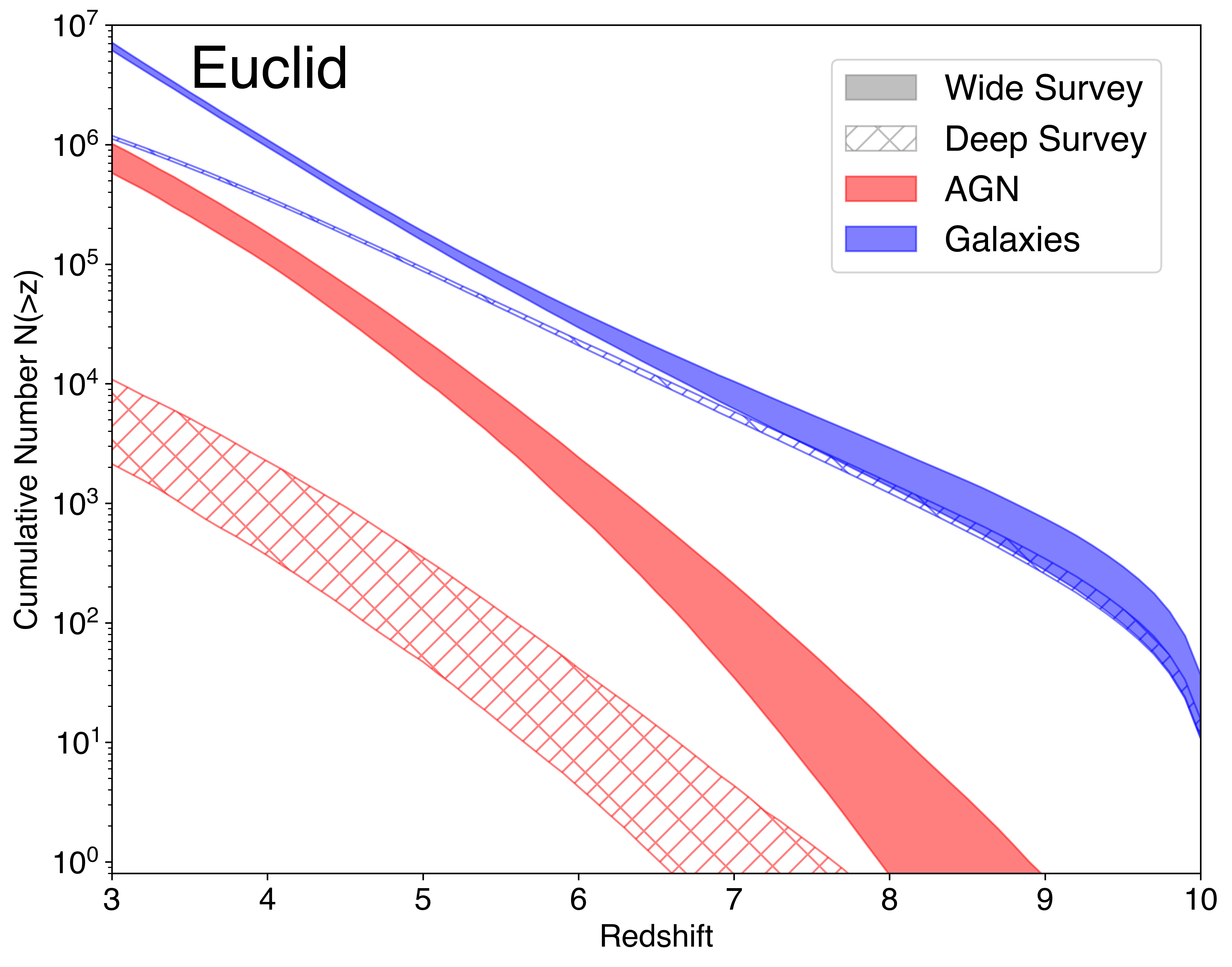}
\plotone{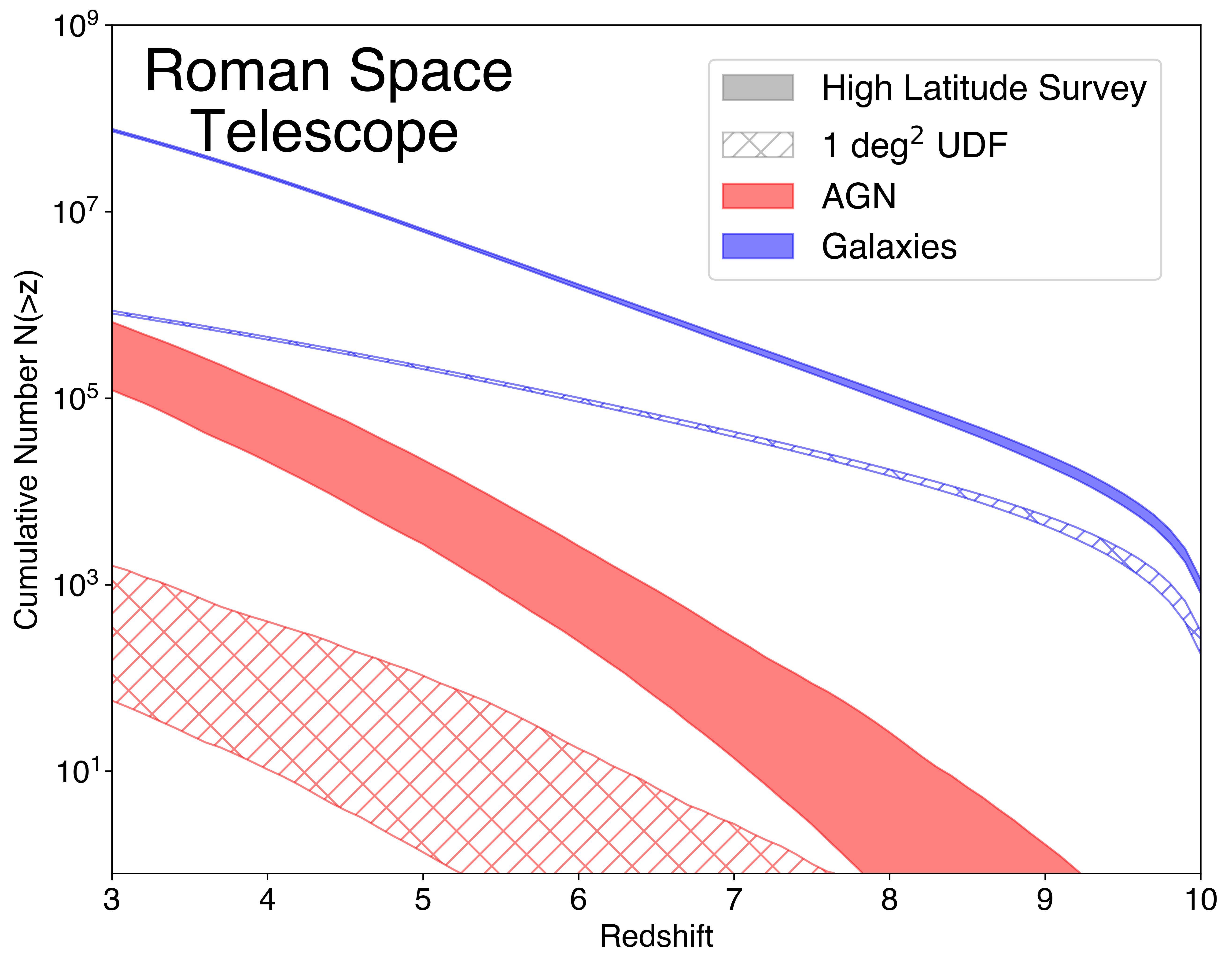}
\vspace*{-3mm}
\caption{Predicted cumulative number (number at redshift $> z$, integrated up to $z =$ 10) of AGN and star-forming galaxies.  The left panel shows the {\it Euclid} Observatory, with the filled and hatched curves denoting the Wide and Deep surveys, respectively.  The right panel shows the predictions for the {\it NGRST}, with the filled and hatched curves denoting the High Latitude Survey and a hypothetical 1 deg$^2$ ultra-deep survey, respectively.  All four surveys will be capable of studying star-forming galaxies out to $z \gtrsim$ 10.  Additionally, we predict that  the {\it Euclid} Wide survey and {\it NGRST} High Latitude Survey will be able to discover large numbers of AGNs at $z >$ 7, constraining our high-redshift AGN luminosity functions predicted here.}
\label{fig:predictions}
\end{figure*}

\begin{deluxetable*}{crrrrrrrr}
\vspace{2mm}
\tabletypesize{\small}
\tablecaption{Predicted Yield from Future Wide-Field Space IR Surveys}
\tablewidth{\textwidth}
\tablehead{
\multicolumn{1}{|c|}{} & \multicolumn{4}{|c|}{{\it Euclid Observatory}} & \multicolumn{4}{|c|}{{\it Nancy Grace Roman Space Telescope}} \\
\hline
\multicolumn{1}{|c|}{} & \multicolumn{2}{|c|}{Star-Forming Galaxies} & \multicolumn{2}{|c|}{AGN} & \multicolumn{2}{|c|}{Star-Forming Galaxies} & \multicolumn{2}{|c|}{AGN}\\
\hline
\multicolumn{1}{|c|}{Redshift} & \multicolumn{1}{|c|}{Wide} & \multicolumn{1}{|c|}{Deep} & \multicolumn{1}{|c|}{Wide} & \multicolumn{1}{|c|}{Deep} & \multicolumn{1}{|c|}{HLS} & \multicolumn{1}{|c|}{UDF} & \multicolumn{1}{|c|}{HLS} & \multicolumn{1}{|c|}{UDF}}
\startdata
$>$3 & 6628979$_{-468791}^{+538774}$ & 1152804$_{-33916}^{+36177}$ & 797214$_{-217778}^{+223326}$ & 5946$_{-3812}^{+4901}$ & 75052128$_{-2149676}^{+2269400}$ & 837215$_{-27360}^{+29259}$ & 356116$_{-234047}^{+300176}$ & 356$_{-299}^{+1241}$ \\
$>$4 & 1025920$_{-69537}^{+73333}$ & 355397$_{-11840}^{+9786}$ & 145278$_{-44302}^{+37428}$ & 1290$_{-926}^{+952}$ & 23904708$_{-745841}^{+651553}$ & 438995$_{-15273}^{+14664}$ & 77530$_{-56716}^{+59656}$ & 78$_{-68}^{+324}$ \\
$>$5 & 172224$_{-15607}^{+16547}$ & 90659$_{-3198}^{+2954}$ & 17108$_{-6209}^{+6767}$ & 155$_{-108}^{+197}$ & 6316084$_{-219632}^{+206850}$ & 212591$_{-8090}^{+7726}$ & 9275$_{-6546}^{+12562}$ & 11$_{-9}^{+95}$ \\
$>$6 & 34695$_{-5146}^{+5841}$ & 22014$_{-1149}^{+1186}$ & 1434$_{-623}^{+989}$ & 14$_{-10}^{+28}$ & 1577002$_{-79258}^{+86396}$ & 95823$_{-4415}^{+4989}$ & 850$_{-604}^{+1752}$ & 1$_{-1}^{+16}$ \\
$>$7 & 8034$_{-1850}^{+2428}$ & 5420$_{-399}^{+428}$ & 88$_{-53}^{+123}$ & 1$_{-1}^{+3}$ & 395302$_{-28133}^{+31234}$ & 40889$_{-2477}^{+2662}$ & 66$_{-52}^{+204}$ & 0$_{-0}^{+3}$ \\
$>$8 & 1988$_{-609}^{+928}$ & 1350$_{-137}^{+139}$ & 4$_{-3}^{+10}$ & --- & 99555$_{-9599}^{+9790}$ & 15814$_{-1150}^{+1390}$ & 3$_{-3}^{+23}$ & --- \\
$>$9 & 445$_{-169}^{+300}$ & 299$_{-43}^{+45}$ & 0$_{-0}^{+1}$ & --- & 22214$_{-3018}^{+2949}$ & 4943$_{-666}^{+572}$ & 0$_{-0}^{+2}$ & --- \\
\enddata
\tablecomments{Predicted cumulative numbers of objects detected for surveys with the {\it Euclid Observatory} and {\it Nancy Grace Roman Space Telescope}.  Numbers were calculated to a magnitude limit 0.5 mag brighter than the stated limit for each survey to allow for robust constraints on the Lyman break, and were integrated up to $z =$ 10.}
\label{tab:predictions}
\end{deluxetable*}

We show these predicted cumulative numbers in Figure~\ref{fig:predictions}, and we list some representative numbers in Table~\ref{tab:predictions}.  For {\it Euclid}, both surveys will find large numbers of star-forming galaxies up to $z \sim$ 9.  While {\it Euclid} may be sensitive to a handful of galaxies at $z \sim$ 10, we have not accounted for the difficulty of selecting such sources with the {\it Euclid} filter suite, as such objects would be detected only in the $H$-band (while {\it NGRST} surveys have the addition of the redder F184W filter).  For AGNs, while the {\it Euclid} Deep survey may find AGNs up to $z \sim$ 7, the Wide survey will find $\sim$100$\times$ more at $z \sim$ 7, and may find AGNs out to $z \gtrsim$ 8.  The results for {\it NGRST} are qualitatively similar.  Both the HLS and the ultra-deep survey will find a large number of sources out to $z \sim$ 10, although in this case the $\sim$2000$\times$ larger area of the HLS results in higher yields at all redshifts.  For AGNs, similar to {\it Euclid} the larger volume probed by the larger survey (the HLS in this case) wins out, with the HLS predicted to find a few AGNs at $z >$ 8, while the ultra-deep survey will not be competitive for very high-redshift AGNs.

The combination of these surveys will result in 1000's of known AGNs at $z >$ 6, and perhaps 10's at $z >$ 8, with each telescope probing different dynamic ranges in luminosity.  These data will be crucial to better constrain the evolution of the AGN luminosity functions deep into the epoch of reionization, improving results such as those presented here.

\subsection{Impact of Double Power Law Assumption}
Here we discuss the impact of our assumption that the star-forming galaxy luminosity function follows a double power law form.  We adopted this as our fiducial model as several recent wide-field surveys find that a DPL is able to better match the observed bright-end number densities when compared to a Schechter function \citep[e.g.,][]{bowler20,harikane21}.  To explore the impact of this assumption, we performed an iteration of our \textsc{emcee} fitting using a Schechter function for the star-forming galaxy luminosity function (still using a DPL for the AGN component). 

To quantitatively distinguish between these two models (DPL$+$DPL and DPL$+$Schechter), we use the deviance information criterion (DIC).  This is a modification of the Bayesian Information criteria
  \citep{liddle04} in that it takes into account both the number of
  data points and the number of free parameters, and it makes use of the full chain.  The DIC is defined as $DIC = -2 (L-P)$, where $L$ is the value of $ln(P)$ of our model using the median of the posterior chains for each parameter, and
$P$ is defined as $P = 2 [L - \frac{1}{N} \sum_{s=1}^s ln P_s]$, where N is the number of samples in the posterior, and $s$ is the sample index. For a model to be preferred over a competing model, it must have a lower DIC.  Here we make use of the updated interpretation by \citet{kass95}, where $\Delta$ DIC $>$ 2/6/10 is positive/strong/decisive evidence against the model with the larger value of DIC.

We calculate the DIC for each model, finding $\Delta_{DIC}$ (defined as the DIC calculated for our fiducial DPL+DPL model minus the DIC for a DPL+Schechter model) of $-$56.  Therefore, our results provide    ``decisive" evidence against the DPL$+$Schechter model.  We further calculate the DIC for each model at each integer redshift of $z =$ 3--9, and find a negative DIC (e.g., in favor of DPL$+$DPL) at all redshifts, implying that it is not the results at any one redshift driving this result.

We do note that the Schechter$+$DPL analysis shows that due to the steeper exponential decline of the galaxy luminosity function, a steeper faint-end slope of the AGN luminosity function would be needed, reaching $\alpha \sim -$3 by $z =$ 8.  This results in a progressively larger fraction of moderately luminous objects required to be AGN at increasing redshifts.  For example, even at $z =$ 9 (7), the AGN fraction reaches 50\% at $M = -$23.5 ($-$23).  This implies a significantly stronger contribution to the global ionizing emissivity from AGNs such that, within the uncertainties, star-forming galaxies and AGNs could contribute equally at $z \sim$ 7.  This scenario seems exceedingly unlikely due to IGM temperature measurements \citep[e.g.][]{hiss18,villasenor21}, as well as the relatively late reionization of \ion{He}{2} \citep[e.g.][]{worseck16}.  While exploring such constraints is beyond this work, the implausibility of this scenario adds additional evidence beyond the DIC results that the star-forming galaxy luminosity function is best modeled by a double power law form.

\section{Conclusions}
We employ a MCMC-based algorithm to jointly model the evolution of the star-forming galaxy and AGN luminosity functions at $z =$ 3--9.  Rather than determine on an object-by-object basis whether an object is dominated by star-formation or AGN activity (difficult in the overlap luminosity regime), we simultaneously fit both luminosity functions to observed volume densities of the entire UV-selected population.  We make use of a variety of published results, including those studying AGNs only, star-forming galaxies only, and the total population.

Our key assumption is that both AGNs and star-forming galaxies have luminosity functions which follow a modified double-power-law form (consisting of a DPL with a faint-end flattening/turnover), and that the parameters which describe this modified DPL evolve smoothly (linearly or quadratically with redshift).  We find that this 29-parameter model is able to successfully fit the large amount of data we consider here.

Our key result, as shown in Figure~\ref{fig:lfratio} is that through this methodology we can self-consistently track the rise in the AGN population from $z =$ 9 to 3.  Between these two redshifts, the onset of AGN activity increased the volume density of UV-luminous objects by $\sim$10$^5$, compared to a rise in the fainter star-forming galaxy population of only $\sim$10$^{2}$.  Consequently, the UV absolute magnitude where AGNs represent 50\% of the population progresses faintward from $M_{UV} \sim -$28.5 at $z =$ 9, to $-$23 at $z =$ 3.

We find that our inferred AGN number density is not high enough to explain the surprisingly high abundance of bright $z =$ 9 galaxies.  Our results imply that a steepening faint-end slope and shallowing bright-end slope towards higher redshift make the star-forming galaxy luminosity function look more and more consistent with a single power law (once observational uncertainties are taken into account).  Both changes are likely driven by reduced feedback from multiple physical processes as one probes earlier epochs in the universe.

We explore the ionizing emissivity inferred from our luminosity functions, and find that star-forming galaxies dominate the intrinsic ionizing emissivity at all redshifts considered. Once plausible escape fractions are taken into account, AGNs begin to play a significant role at $z <$ 4, but they are still sub-dominant to galaxies even at $z =$ 3.  We find that our inferred AGN luminosity functions are not inconsistent with several theoretical predictions which employ a variety of black hole seeding and growth models.  However, all predictions we considered failed to reproduce the observed abundance of bright AGNs.

Finally, we use our inferred luminosity functions to make predictions for the next generation of wide-field space-based near-infrared observatories.  We explore both the Wide and Deep {\it Euclid Observatory} surveys, and the High Latitude Survey and a hypothetical 1 deg$^2$ ultra-deep survey from the {\it Nancy Grace Roman Space Telescope}.  While all four surveys will enable the discovery of large numbers of star-forming galaxies out to $z \sim$ 10, only the {\it Euclid} Wide and {\it NGRST} HLS will obtain significant numbers of AGNs deep into the epoch of reionization.

Our methodology shows the discovery power possible when combining a large amount of data over a wide dynamic range in luminosity.  However, there are degeneracies between the various parameters used, thus these results can be validated in a variety of ways. First, spectroscopic followup of sources in the AGN/galaxy overlap region can empirically measure the AGN fraction.  Second, the nature of the highest redshift sources used to constrain our luminosity functions are still poorly known, thus spectroscopic confirmation of their redshifts will increase confidence.  Finally, direct measurement of the AGN luminosity function out to $z \geq$ 7 with {\it Euclid} and {\it NGRST} will directly test our predicted AGN luminosity function evolution.  The impending flow of data from {\it JWST} this year, and {\it Euclid}, {\it NGRST} and the {\it Vera Rubin Observatory} later this decade will provide the needed observations to better constrain predictions such as those we have presented here.

\begin{acknowledgements}
We acknowledge that the location where this work took place, the University of Texas at Austin, that sits on indigenous land. The Tonkawa lived in central Texas and the Comanche and Apache moved through this area. We pay our respects to all the American Indian and Indigenous Peoples and
communities who have been or have become a part of these lands and territories in Texas, on this piece of Turtle Island. 

SLF thanks the dozens of colleagues who have berated him over the years to learn Python, forcing him to spend his sabbatical learning this new language, which led to this project.    We thank Rachel Somerville, Aaron Yung, Anson D'Aloisio, Priya Natarajan, Casey Papovich, Russell Ryan, Junehoyung Jeon and Volker Bromm for stimulating scientific conversations which improved the discussion in this paper. We thank Girish Kulkarni, Angelo Ricarte, Yuichi Harikane, Thibaud Moutard, Shotaro Kikuchihara, Masami Ouchi, Junehyoung Jeon, Aaron Yung and Linhua Jiang for sharing their data with us.  SLF and MB acknowledge support from NASA through ADAP award 80NSSC18K0954. SLF acnowledges support from the University of Texas at Austin Faculty Research Assignment program.
\end{acknowledgements}

\bibliographystyle{apj}
\bibliography{stevenf}


\end{document}